\documentclass[oneside]{article}
\usepackage[T1]{fontenc}
\usepackage{authblk}
\usepackage{float}
\usepackage{amsmath}
\usepackage{graphicx}
\usepackage{xfrac}
\usepackage[english]{babel}
\usepackage{lmodern}
\usepackage[T1]{fontenc}
\usepackage[latin9]{inputenc}

\usepackage{csquotes}

\usepackage{mathtools}
\usepackage{graphicx}
\usepackage{esint}
\usepackage[unicode=true,
 bookmarks=false,
breaklinks=false,pdfborder={0 0 1},backref=false,colorlinks=false]
 {hyperref}
\usepackage[a4paper, total={210mm,297mm},margin=2.0cm]{geometry}
\usepackage[usenames,dvipsnames]{xcolor}
\usepackage{subcaption}
 \usepackage{enumerate}

\usepackage{newfloat}
\DeclareFloatingEnvironment[name={Supplementary Figure}]{suppfigure}
\DeclareFloatingEnvironment[name={Supplementary Table}]{supptable}
\title{Particle Shape Influences  Settling and  Sorting Behavior in Microfluidic Domains}

\author[1]{Hakan Ba\c sa\u gao\u glu \thanks{Electronic address: \texttt{hbasagaoglu@swri.org}; Corresponding author}}
\author[2]{Sauro Succi}
\author[3]{Danielle Wyrick}
\author[4]{Justin Blount}

\affil[1]{Mechanical Engineering Division, Southwest Research Institute, San Antonio, TX 78238 USA}
\affil[2]{Istituto Applicazioni del Calcolo, via dei taurini 19, 00185, Roma, Italy }
\affil[3]{Space Science Division, Southwest Research Institute, San Antonio, TX 78238 USA}
\affil[4]{Defense Intelligence Solutions Division, Southwest Research Institute, San Antonio, TX 78238 USA}

\date{\today}

\begin{document}

\maketitle

\begin{abstract}
We present a new numerical model to simulate settling  trajectories of discretized individual or a mixture of particles of different geometrical shapes in a quiescent fluid and their flow trajectories in a flowing fluid. Simulations unveiled diverse particle settling trajectories as a function of their geometrical shape and density. The effects of the surface concavity of a boomerang particle and aspect ratio of a rectangular particle on the periodicity and amplitude of oscillations in their settling trajectories were numerically captured. Use of surrogate circular particles for settling or flowing of a mixture of non-circular particles were shown to miscalculate particle velocities by a factor of 0.9-2.2 and inaccurately determine the particles' trajectories. In a microfluidic chamber with particles of different shapes and sizes, simulations showed that steady vortices do not necessarily always control particle entrapments, nor do larger particles get selectively and consistently entrapped in steady vortices. Strikingly, a change in the shape of large particles from circular to elliptical resulted in stronger entrapments of smaller circular particles, but enhanced outflows of larger particles, which could be an alternative microfluidics-based method for sorting and separation of particles of different sizes and shapes. 
\end{abstract}

\section*{Introduction}
Flow and transport of engineered particles of different geometrical shapes are encountered in diverse biomedical applications. In targeted drug deliveries, the shape of engineered drug cargos has shown to have intriguing effects on their transport in blood vessels, adhesion onto channel walls, and targeting ability toward malignant cells \cite{CKM07}. For example, ellipsoidal microparticles displayed longer blood circulation times than spherical particles due to less efficient phagocytosis by macrophages in the reticuloendothelial system \cite{SVA10}. Hexagonal nanoparticles more effectively mitigated phagocytoses and remained in blood circulation longer than spherical particles \cite{LHL11}. Unlike spherical particles, boomerang-shaped particles displayed a preferred direction of Brownian motion \cite{CKW13}, which could have implications in design of new microscopic particles to deliver drugs or self-assemble into complex materials. A theranostic plasmonic shell-magnetic core star-shaped nanomaterial was used for targeted isolation and detection of rare tumor cells from a blood sample \cite{FSS12}. As for the adhesion kinetics of such engineered particles on channel walls, nanorod particles were numerically shown to adhere to channel walls easier than spherical particles due, in part, to larger surface area contacts with the channel walls as they tumble near the walls \cite{SLHG11}. 

In applications relevant to the design of biomedical devices, microfluidic devices with different geometric designs have been proposed to isolate circulating tumor cells (CTC) from healthy cells in blood samples through, for example, vortex-aided particle separation \cite{PCD17,ZKP13}, which could be useful for early cancer diagnosis and monitoring metastatic progression or the efficiency of cancer treatments \cite{AP13}. Although the performance of the microfluidic devices in the segregation of CTC has been commonly tested with surrogate spherical particles, tumor cells often exhibit patient-specific arbitrary shape profiles, which do not conform to the spherical particle representation for tumor cells \cite{PAD14,MBL14}. 

The effect of non-spherical particle shapes on particle trajectories has been recently addressed in numerical simulations. Settling dynamics and patterns of thin disks\cite{AMF13,CBD13} in an infinitely long viscous fluid domain and settling behaviors of  individual spherical, cubical, or tetrahedral particles in an infinitely long fluidic domain with periodic lateral boundaries\cite{RW14} were numerically investigated. However, numerical simulations of settling of a \textit{mixture} of different-shaped particles (DSP), involving angular- and curved-shaped particles, in a bounded domain is unprecedented. Similarly, numerical simulations of flow trajectories of a \textit{mixture} of DSP is very limited or perhaps non-existent in the literature.      

The extension of the lattice Boltzmann (LB) method for simulating flow of suspended bodies is a fast-growing area of LB research \cite{S15}, following the pioneering work of of Ladd \cite{L94a,L94b}. Considering broad uses of DSP in biomedical applications and the abundant experimental evidence for their shape-dependent distinct flow and transport behaviors, we extended the LB model (LBM) presented originally by Nguyen and Ladd\cite{NL02} to simulate the settling and flow of DSP, including discretized angular-shaped particles (DAsP), involving star, boomerang, hexagonal, triangular, rectangular, and discretized curved-shaped particle (DCsP), involving circular and elliptical particles, consistent with the aforementioned shapes of engineered particles used in biomedical applications. The DSP-LBM is suitable for simulating settling and flow trajectories of any arbitrary-shaped particles, such as tumor cells. 

The primary purpose of this paper is to introduce the  DSP-LBM and demonstrate its performance in simulating the settling or flow of individual or a mixture of DSP under various combinations of properties associated with the particles, flow regimes, and the microfluidic domain geometry. Using the DSP-LBM and a single chamber of the microfluidic device geometry in Ref \cite{PCD17}, we numerically investigated the validity of recent findings and implications in microfluidic research.  These findings and implications involve: (i) when a large number of particles are released into a fluid in a microfluidic device, larger particles get selectively trapped by vortices, whereas smaller particles avoid  entrapments; (ii) steady vortex structures can be used to quantify vortex-controlled, size-based separation of particles; and (iii) non-circular particles may be represented by circular particles in vortex-aided particle segregation via microfluidic devices with different geometric peculiarities.   

\section*{Methods}
In the LB method\cite{HS89,BSV92,S01,W00}, the mesodynamics of the Newtonian fluid flow can be described by a single relaxation time via the  Bhatnagar-Gross-Krook (BKG) equation \cite{BGK54}

\begin{equation}
 \label{e.LB1} f_{i}\left(\mathbf{r+e}_{i}{\triangle t},t+{\triangle
 t}\right) -f_{i}\left(\mathbf{r},t \right) =\frac{\triangle t}{\tau} [
 {f_{i}^{eq}\left(\mathbf{r},t \right)-f_{i}\left(\mathbf{r},t
 \right) } ],
 \end{equation}

\noindent where $f_i(\mathbf{r},t)$ is the set of population densities of discrete velocities $\mathbf{e}_i$ at position $\bf{r}$ and discrete time $t$ with a time increment of $\triangle t$, $\tau$ is the relaxation parameter, and $f_i^{eq}$ is the local equilibrium \cite{QDL92}, $f_{i}^{eq}=\omega_i \rho 
\left
 [1+\left(\mathbf{e}_i\mathbf{\cdot}\mathbf{u} \right)/{c_s^2}
 +\left(\mathbf{e}_i\mathbf{\cdot}\mathbf{u})^2 \right) /{2c_s^4}-
 \left(\mathbf{u \mathbf{\cdot}u}\right) / {2c_s^2}\right] $,  $\omega_i$ is the weight associated with $\mathbf{e}_i$ and $c_s$ is the speed of sound, $c_s= \triangle x /\sqrt(3) \triangle t$. The local fluid density, $\rho$, and velocity, $\mathbf{u}$, at the lattice node are given by $\rho=\sum_{i} f_{i}$ and $\rho \mathbf{u}=\sum_{i} f_{i}\mathbf{e}_i+\tau \rho \mathbf{g}$, where $\mathbf{g}$ is the strength of an external force \cite{BG00}. A D2Q9 (two-dimensional nine velocity vector) lattice \cite{S01} was adopted in numerical simulations. Through the Chapman-Enskog approach, the LB method for a single-phase flow recovers the Navier-Stokes equation in the limit of small Knudsen number for weakly compressible fluids, in which $\nabla \cdot \mathbf{u} \sim 0$ and $\partial_{t} \mathbf{u} +\left( \mathbf{u} \cdot \nabla \right) \mathbf{u} = -\left( \nabla P / {\rho} \right) + \nu \nabla^2 \mathbf{u} + \mathbf{g}$ with the fluid kinematic viscosity, $\nu= c_s^2 \triangle t \left( \tau - 0.5 \right) $. Pressure, $P$, is computed via the ideal gas relation, $P=c_s^2 \rho$.  

The extension of the LBM to the DSP-LBM involves (i) geometric description of DSP to locate the vertices for DAsP or boundary nodes for DCsP, (ii) calculations of the position of intra- and extra-particle boundary nodes in the vicinity of arbitrary-shaped particle surfaces across which the particle and fluid exchange momentum, and (iii) calculations of new positions of the center of mass of a particle and its vertices based on particle-fluid hydrodynamics.

\subsection*{Geometric Description of 2D Different-shaped Particles}\label{asp}
\begin{figure}[h!]
\begin{center}
\scalebox{0.39} {\includegraphics{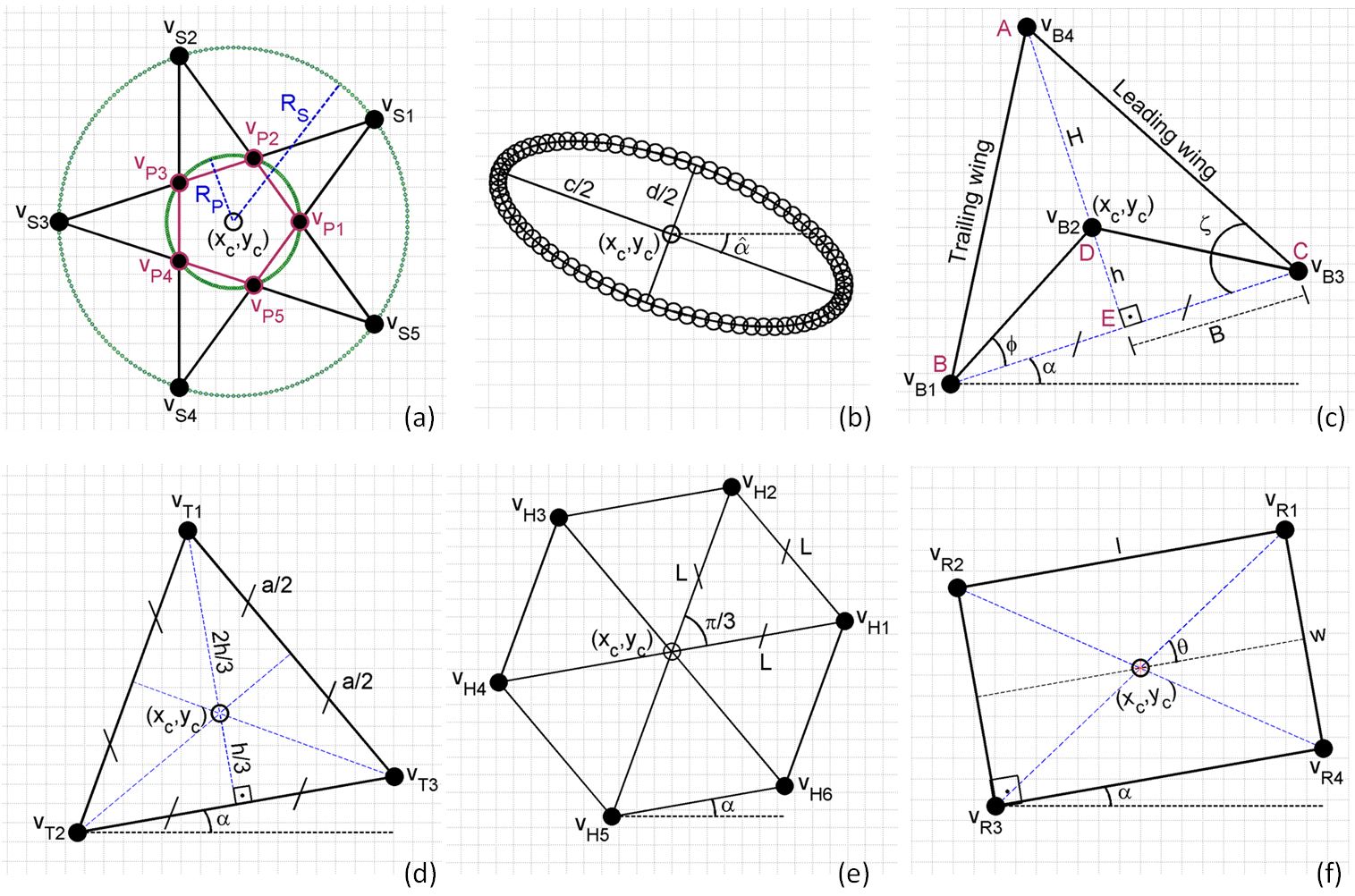}} 
\caption{ A schematic representation of non-circular particle geometries in the DSP-LBM. $\hat\alpha > 0^{\circ}$ and $\alpha > 0^{\circ}$ represent the initial tilt angle in the clockwise and counterclockwise directions. $\hat\alpha = 0^{\circ}$ in (a). \label{Fig:Geometry}}\label{fig:Geometries}
\end{center}
\end{figure}

Similar to geometric construction of surfaces of a circular-cylindrical particle (hereafter, circular particle) by Ladd\cite{NL02}, we used discretized particle surfaces for 2D curved (e.g., circular)- and angular (e.g., hexagonal)-shaped particles in the DSP-LBM. A schematic representation of non-circular particle geometries are shown in Fig. \ref{fig:Geometries}, which are subsequently used to locate vertices of DAsP and boundary nodes of DCsP. We provide geometric descriptions for the star-shaped and elliptical particles next, but geometric descriptions of the remaining particles are provided in Supplementary Information-1.

The star-shaped particle geometry is represented by five isosceles triangles connected to a pentagon at the center, as shown in Fig. \ref{fig:Geometries}a. The geometry is constructed by two circles; the bigger circle with a radius of $R_S$ encloses the star-shape and the smaller circle with a radius of $R_P$ that passes through the corners of the pentagon.  These two circles are related via $R_S=\psi R_P$, in which $\psi=cos \left(\pi/5 \right)+\left[ sin \left(\pi/5 \right)\right] / \left[tan \left(\pi/10 \right)\right] $.   The surface area of the star-shaped particle, $A_S$, is given by $A_S=\chi \left( R_S \right)^2$, in which  $\chi= \left[ sin^2 \left(  \pi/5  \right)  /  {\psi^2} \right]  \left[  {5} / {tan\left( \pi/10\right]}  +4\psi \right)$. The star-shaped particle has five vertices located on the outermost tip of the triangles ($v_{S1} - v_{S5}$), in addition to five vertices located on the corners of the inner pentagon ($v_{P1} - v_{P5}$) (Fig. \ref{fig:Geometries}a).  The coordinates $\left(x_i, y_i\right)$ of $v_{Si}$, and $v_{Pi}$, where $i \epsilon \left[1,5\right]$, are computed by Eq. \ref{star_e1} and Eq. \ref{star_e2}, respectively,

\begin{equation} 
  \label{star_e1} 
 \left[ \begin{array}{c} x_i\\ y_i \end{array} \right] = \left[ \begin{array}{c} x_c \\ y_c \end{array} \right]+  R_S  \left[ \begin{array}{c} cos(\hat{\alpha}+\left(2i-1\right) \pi/5) \\  sin(\hat{\alpha}+\left(2i-1\right) \pi/5) \end{array} \right], 
 \end{equation}

\vspace*{0.01cm} 
 
 \begin{equation} 
  \label{star_e2} 
 \left[ \begin{array}{c} x_i\\ y_i \end{array} \right] = \left[ \begin{array}{c} x_c \\ y_c \end{array} \right]+  R_P \left[ \begin{array}{c} cos(\hat{\alpha}+2\left(i-1\right) \pi/5) \\ sin(\hat{\alpha}+2\left(i-1\right) \pi/5) \end{array} \right], 
 \end{equation}

\noindent
where $\hat{\alpha}$ is the initial tilt angle of the particle in the clockwise direction. $\mathbf{x}_c=\left(x_c, y_c \right)$ is the center of mass of a particle, $x_c= \frac{1}{N} \sum_{i=1}^N {x_i}$ and $y_c=\frac{1}{N} \sum_{i=1}^N {y_i}$, where $N$ is the number of vertices ($N_{Ver}$) for DAsP or the number of boundary nodes ($N_{Bnd}$) for DCsP. The mass of the star-shaped particle per unit particle thickness is given by $m_p= \chi \left( R_S \right)^2 \rho_p$. The moment of inertia, $I_s$, for the star-shaped particle was computed by $I_s= \left( {A_P \rho_p a^2}/{24} \right)   \left[ 1+3cot^2 \left(  \frac{\pi}{5}  \right)     \right] + \left( {5A_T \rho_p}/{72} \right) \left(  4h^2+3a^2  \right) +2 A_T \left( \lambda +h/3 \right)^2 \sigma $, in which $a$ is the side length of the pentagon, $a=2R_Psin\left( \pi/5\right)$, $h$ is the height of an isosceles triangle,   $h=a/\left[2 tan \left(\pi/10\right) \right]$, $A_P$ is the area of the pentagon,  $A_P=\left( 1/4\right) \sqrt{5+\left( 5+2  \sqrt{5}\right)}a^2$, $A_T$ is the area of the triangle, $A_T=ah/2$, $\lambda=R_p cos(\pi/5)$, and $\sigma=\left[cos(\pi/5)+cos(2\pi/5)\right]^2+\left[0.5+sin(\pi/5)+sin(2\pi/5)\right]^2$.

Different from a star-shaped particle, the elliptical particle geometry is described by boundary nodes, $N_{bnd}$, along the discretized curved surfaces, the length of its long- and short-axes ($c$ and $d$), and the initial tilt angle, $\hat{\alpha}$ (Fig. \ref{fig:Geometries}b). The coordinates of its boundary nodes are computed by

 \begin{equation} 
  \label{asp_e3} 
 \left[ \begin{array}{c} x_i\\ y_i \end{array} \right] = \left[ \begin{array}{c} x_c \\ y_c \end{array} \right]+  \begin{bmatrix} cos(\Phi_i) cos(\hat{\alpha}) & - sin (\Phi_i)  sin(\hat{\alpha}) \\ cos(\Phi_i) sin(\hat{\alpha})  &  sin (\Phi_i)  cos(\hat\alpha) \end{bmatrix} \left[ \begin{array}{c} c/2 \\ d/2 \end{array} \right] 
 \end{equation}

\noindent
in which $\Phi_i=2\pi \left( i-1 \right) / {\left( N_{Nbd}-1 \right)}$. The mass of an elliptical particle per unit particle thickness is given by $m_p=A_E \rho_p$, in which the surface area and its moment of inertia are computed by $A_E=\pi cd/4$ and $I_E = \frac{m}{16} \left( c^2 + d^2 \right)$, respectively. 

\subsection*{Intra-Particle Boundary Nodes (IPBN) and Extra-Particle Boundary Nodes (EPBN)}
\begin{figure}[h!]
\begin{center}
\scalebox{0.44} {\includegraphics{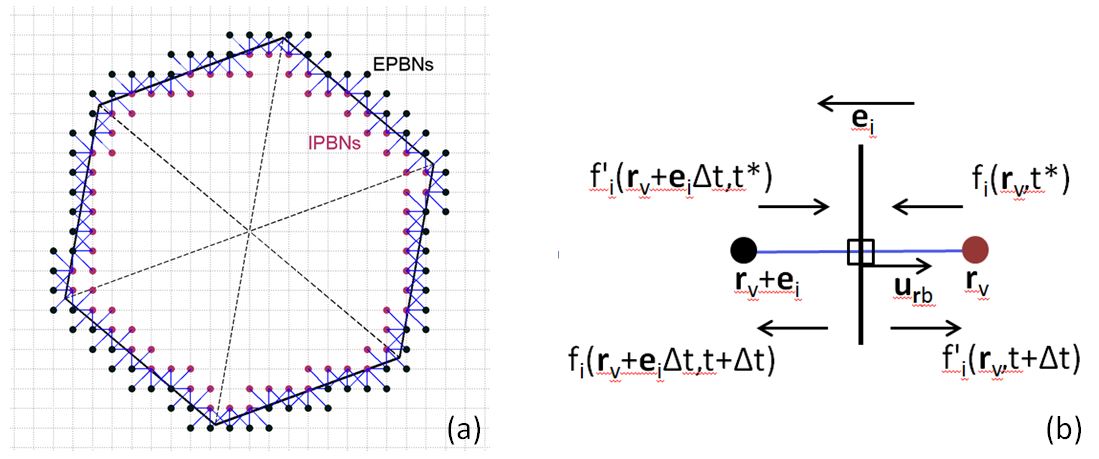}} 
\caption{(a) IPBNs and EPBNs of a discretized hexagonal particle geometry in the DPS-LBM. Blue lines are the hydrodynamic links along which the particle and fluid exchange momentum. (b) Momentum exchange between the particle and fluid at a boundary node marked by a square. \label{Fig:Geometry}}\label{Fig:FluidParHyd}
\end{center}
\end{figure}

The winding number algorithm\cite{O98} was implemented to determine whether a lattice node $\mathbf{x}_k=\left(x_k,y_k \right) $ is enclosed by a polygon in Fig. \ref{fig:Geometries}, described by a series of boundary nodes for DCaP or vertices for DAsP along the particle surface. The algorithm computes the number of times the polygon winds around $\mathbf{x}_k$, which is referred to as the winding number, $m \left(\mathbf{x}_k \right) $.  $\mathbf{x}_k$ is not enclosed by a polygon if $m \left(\mathbf{x}_k \right)=0$. In the DSP-LBM, $\mathbf{x}_k$ and $\mathbf{x}_k+\mathbf{e_i}$ form a intra-particle boundary nodes (IPBN)  and extra-particle boundary nodes (EPBN) pair if $m \left( \mathbf{x}_k \right) \neq 0$ and $m \left( \mathbf{x_k+e_i} \right) = 0$. The IPBNs and EPBNs for a discretized hexagonal particle and the momentum exchanges between the particle and the fluid at the mid-point of hydrodynamic links connecting an IPBN and an EPBN are shown in Fig. \ref{Fig:FluidParHyd}.

\subsection*{Particle-fluid Hydrodynamics}
\label{sec:pfh}
\definecolor{codepurple}{rgb}{0.58,0,0.82}
Particle-fluid hydrodynamic calculations rely on momentum exchanges between the fluid and the mobile DSP, following the approach in Ref. \cite{NL02,BS10}, in which the population densities near particle surfaces are modified to account for momentum-conserving particle-fluid collisions. Particle-fluid hydrodynamic forces, $\mathbf{F}_{\mathbf{r}_b}$, at the boundary nodes located halfway between the intra-particle lattice node, $\mathbf{r}_v$, and extra-particle lattice node, $\mathbf{r}_v + \mathbf{e}_i$, are computed by \cite{L94a,DL03,BCF15} 

 \begin{equation}
 \label{e.pfh3} \mathbf{F}_{\mathbf{r}_b}=-2\left[f^{\prime}_i\left(\mathbf{r}_v+\mathbf{e}_i \triangle t,t^{\ast}\right)+
\frac{\rho\omega_i}{c_s^2}\left(\mathbf{u}_{\mathbf{r}_b} \cdot
\mathbf{e}_i\right)\right]\mathbf{e}_i.
 \end{equation}

\noindent The translational velocity, $\mathbf{U}_p$, and the angular velocity of the particle, $\Omega_p$, are advanced in time according to the discretized Newton's equations of motion,  
$\mathbf{U}_p\left( t+\triangle t\right) \equiv \mathbf{U}_p\left(t\right)+\frac{\triangle t}{m_p}\mathbf{F}_T \left( t\right)+\frac{\triangle t}{\rho_p}(\rho_p-\rho)\mathbf{g}$ and $\Omega_p\left(t+\triangle t\right) \equiv \Omega_p\left(t\right)+\frac{\triangle t}{I_p}\mathbf{T}_T\left( t\right)$, where $m_p$ is the particle mass, $I_p$ is the moment of inertia of the particle, and $\mathbf{u}_b=\mathbf{U}_p+\Omega_p\times\left({\mathbf{r}_b} 
- \mathbf{r}_c \right)$. The new position of the center of mass of a particle is computed as $\mathbf{x}_c\left(t+\triangle t \right) =\mathbf{x}_c \left( t\right)+\mathbf{U}_p\left(t \right)\triangle
t$. The population densities at $\mathbf{r}_v$ and $\mathbf{r}_v +\mathbf{e}_i \triangle t$ are updated to account for particle-fluid hydrodynamics in accordance with \cite{L94a}

\begin{equation}
 \label{e.MP7} f^{\prime}_i\left(\mathbf{r}_v,t+\triangle t\right)=f_i(\mathbf{r}_v,t^{\ast})-\frac{2\rho \omega_i}{c_s^2}\left(\mathbf{u}_{\mathbf{r}_b} \cdot \mathbf{e}_i \right), \\
 f_i\left(\mathbf{r}_v+\mathbf{e}_i \triangle t,t+\triangle t\right)=f^{\prime}_i(\mathbf{r}_v+\mathbf{e}_i \triangle t,t^{\ast})+\frac{2\rho
\omega_i}{c_s^2}\left(\mathbf{u}_{\mathbf{r}_b}  \cdot \mathbf{e}_i \right).
 \end{equation}
\subsection*{New Locations of Vertices or Boundary Nodes}
The locations of vertices or boundary nodes are updated in each time step. The distance $\mathbf{d}_i=\left(d_{ix},d_{iy} \right)$ between the $i^{th}$ vertex (or a boundary node) and the center of mass of a particle, $\mathbf{x}_c$ is computed via $\mathbf{d_i}=\mathbf{x}_i - \mathbf{x_c} $. After $\mathbf{x}_c\left(t+\triangle t \right)$ is computed, new positions of vertices (or boundary nodes) are updated via
\begin{equation} 
  \label{star_e3} 
 \left[ \begin{array}{c} x_i\\ y_i \end{array} \right] = \left[ \begin{array}{c} x_c \\ y_c \end{array} \right]+   \left[ \begin{array}{c} d_{ix} cos\left( \left(\Omega_p+\Upsilon_i \right) \triangle t\right) \\ d_{iy} sin \left( \left(\Omega_p+\Upsilon_i \right) \triangle t\right) \end{array} \right] 
 \end{equation}

\noindent in which $\Upsilon_i$ is the angle between $\left( \mathbf{x}_i -\mathbf{x_c} \right) $ and $+x$. For a hexagonal particle, for example, $\Upsilon_i=\alpha+(i-1) \pi / {3}$ for $i\epsilon[1,6]$.


\subsection*{Model Validation}

\begin{figure}[h!]
\begin{center}
\scalebox{0.45} {\includegraphics{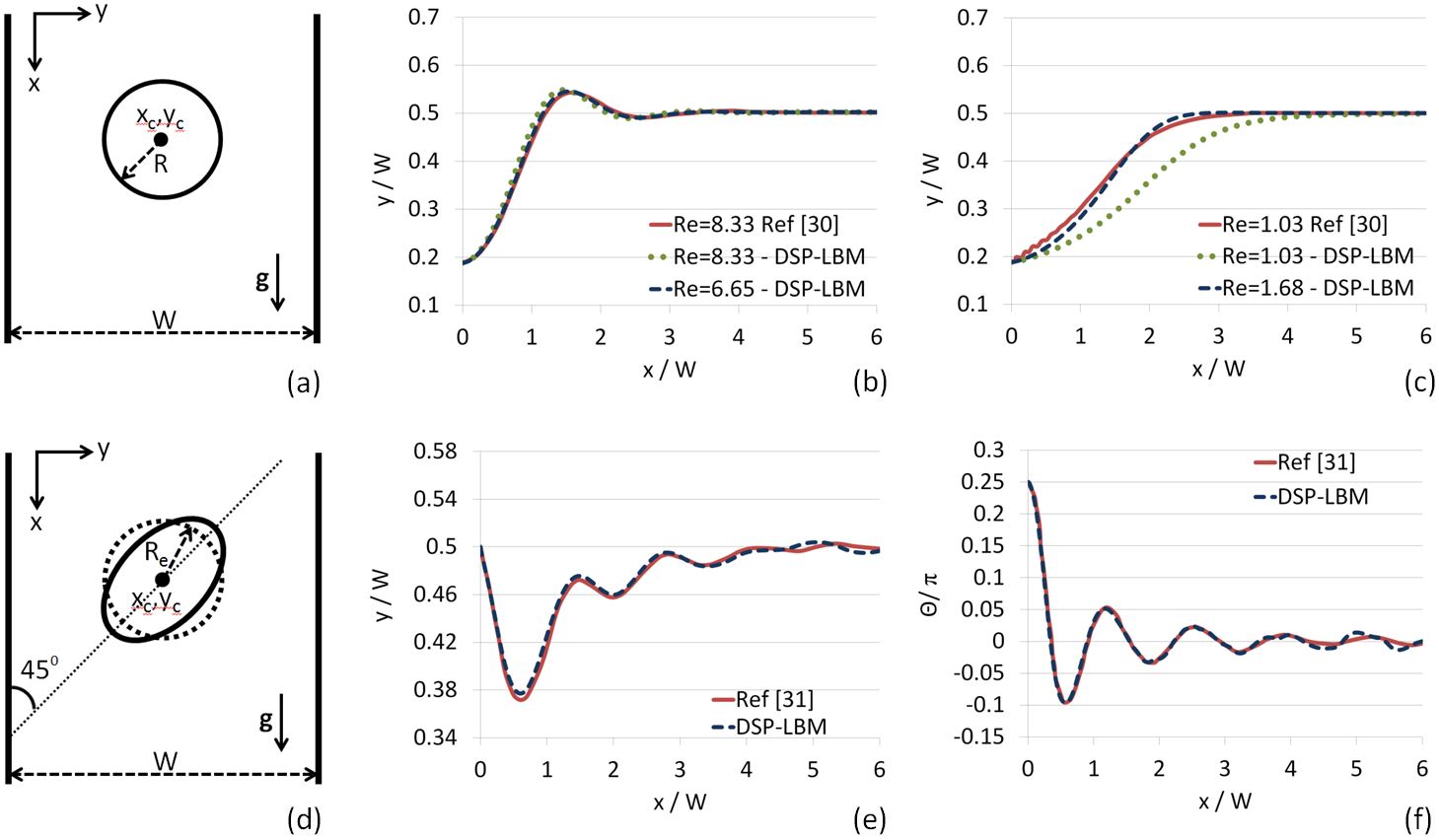}} 
\caption{Numerical validations of the DSP-LBM with two benchmark problems, involving settling of a circular particle in (a)-(c) and an elliptical particle in (d)-(f).   \label{Fig:VT_DSP_LBM}}
\end{center}
\end{figure}

The DSP-LBM was validated with two benchmark problems. First, the settling trajectory of a circular particle in an initially quiescent fluid in a bounded domain (Fig. \ref{Fig:VT_DSP_LBM}a) computed by the DSP-LBM was compared against the finite-element (FE) solutions by Feng \textit{et al.} \cite{FHJ94} at two different  Reynolds numbers, $Re=8.33$ and $Re=1.03$ (here, $R_e=2R U_s/\nu$, where $R$ is the particle radius and $U_s$ is the settling (terminal) velocity of the particle). In Ref. \cite{FHJ94}, the values of $R$ and $\nu$ in FE simulations were not provided, but only $Re$ values were reported. In the DSP-LBM simulations, the length of the bounded flow domain was set to $\sim30W$ (adopted in all settling simulations in this paper), where $W$ is the channel width perpendicular to the main settling direction, and $R$=385 $\mu$m, $\nu=0.01$ cm$^2$, and $|g|=981$ cm$/$s$^2$. $\rho_p/\rho$ was adjusted to meet the reported $Re$ values in Ref. \cite{FHJ94}. For $Re=8.33$, DSP-LBM (with $\rho_p/\rho$=1.07) and FE solutions are in good agreement (Fig. \ref{Fig:VT_DSP_LBM}b), although the DSP-LBM solution for $Re=6.65$ (with $\rho_p/\rho=1.05$) matched the FE solution for $Re=8.33$ better. The FE solution for $Re=1.03$ was in a good agreement with the DSP-LBM solution (with $\rho_p/\rho=1.01$) for $Re=1.68$ (Fig. \ref{Fig:VT_DSP_LBM}c). 

In the second validation test, the DSP-LBM simulation of the settling trajectory and angular rotations $\left( \theta=\hat\alpha+\Omega_p \triangle t \right)$ of an elliptical particle in an initially quiescent fluid in a bounded domain (Fig. \ref{Fig:VT_DSP_LBM}d) was compared against numerical solutions by Xia \textit{et al.}\cite{ZCR09}. In these simulations, $c/d=2$, $W/c=4$ (Fig. \ref{Fig:VT_DSP_LBM}d), density ratio of $\rho_p / \rho=1.1$, $\hat{\alpha}=45^{\circ}$, $\nu=0.01$ cm$^2 /$s, $c=0.1$ cm, and $|\mathbf{g}|=981$ cm$/$s$^2$. Figs. \ref{Fig:VT_DSP_LBM}e and \ref{Fig:VT_DSP_LBM}f show that settling trajectory and angular rotations of an elliptical particle computed by DSP-LBM are in good agreement with the simulation results by Xia \textit{et al.} \cite{ZCR09}
 
 \subsection*{Data Availability}
The datasets generated during and/or analysed during the current study are available from the corresponding author on reasonable request.

\section*{Results}

\textbf{Settling Trajectories of Different-Shaped Particles}

The DSP-LBM was used to simulate the settling trajectories and velocities of DSP as a function of particle density. The same problem set-up in Fig. \ref{Fig:VT_DSP_LBM}d was used, but the elliptical particle was replaced by particles of different shapes. The blockage ratio is defined as $W/R_e$, in which $R_e$ is the equivalent radius of a circular particle that has the same surface area of a non-circular particle. In these simulations, $R_{e}=3.5\times10^{-2}$ cm, the surface area of the particle is $A_p=3.9\times10^{-3}$ cm$^2$, and $g$=981 cm/s$^2$. 

\begin{figure}[h!]
\begin{center}
\scalebox{0.44} {\includegraphics{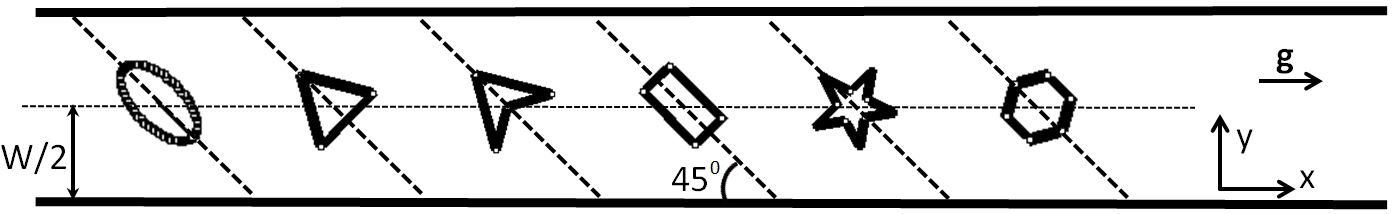}} 
\caption{A schematic representation of initial orientations of non-circular particles in the settling simulation. The center of mass of the particles was initially located on the mid-channel ($y=W/2)$ near the inlet.  \label{Fig:Initial_Orientation}}
\end{center}
\end{figure}

The same $A_p$ was specified for all DSP by setting $R_s=15.4$, and $R_p=5.9$ for the star particle; $B=60$, $\zeta=\pi/3$, $\phi=\pi/6$ for the boomerang particle; $L=10.1$ for the hexagonal particle; $a=24.8$ for the triangular particle; $l=23.1$, $w=11.5$ for the rectangular particle; $R=R_e=9.2$ for the circular particle; and $c=26$, $d=13$ for the elliptical particle (Fig. \ref{fig:Geometries}). Here, the length parameters are expressed in l.u. (1 l.u. $=3.846\times10^{-3}$ cm) and angles are described in radians. The initial orientation of the particles are shown in Fig. \ref{Fig:Initial_Orientation}.

\begin{figure}[h!]
\begin{center}
\scalebox{0.44} {\includegraphics{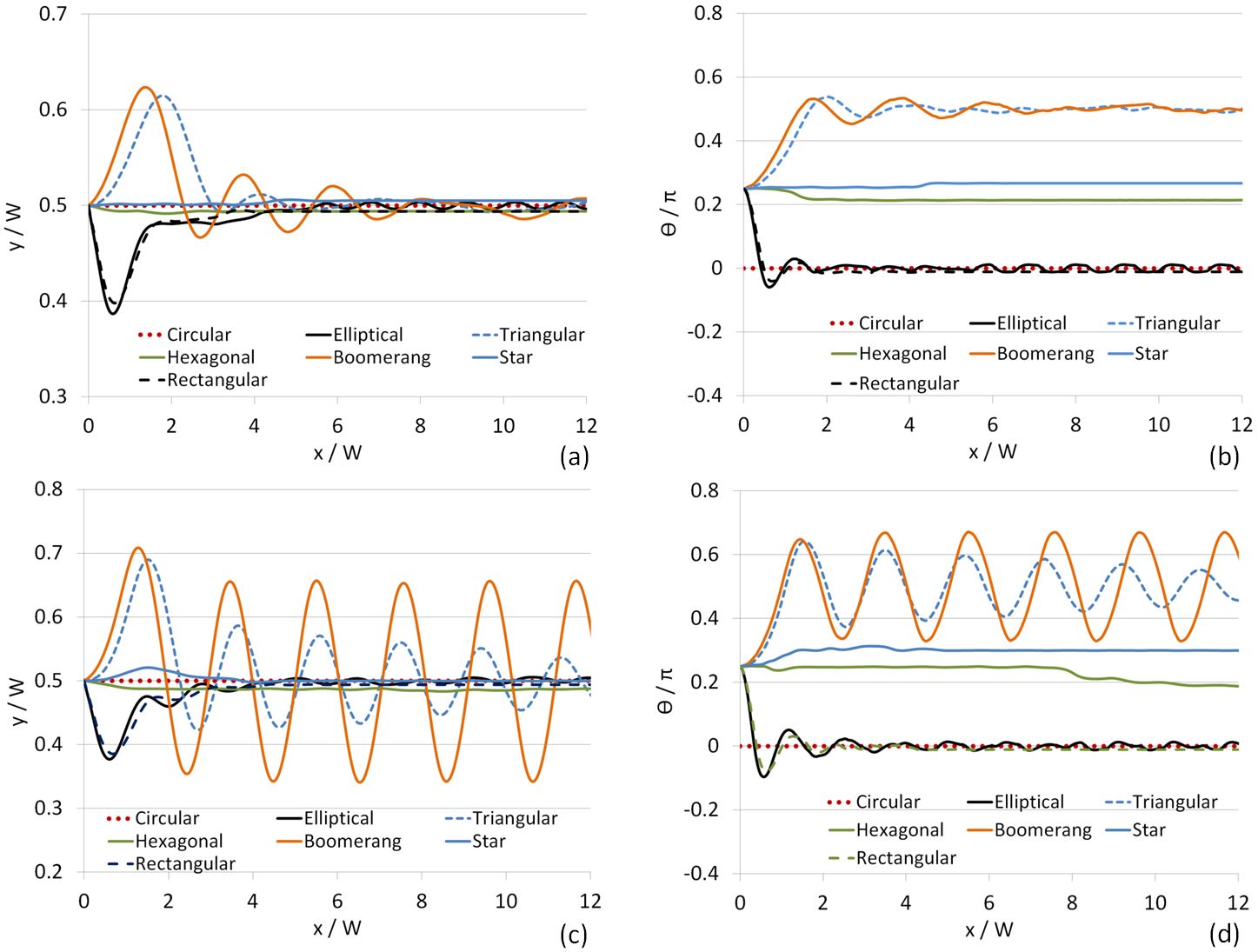}} 
\caption{Settling trajectories and angular rotations of DSP with $\rho / \rho_p$ =1.05 (a-b) and with  $\rho / \rho_p$ =1.10 (c-d). $W/R_e=11.3$ in these simulations.   \label{Fig:Settling_BC}}
\end{center}
\end{figure}

DSP-LBM simulation results in Fig. \ref{Fig:Settling_BC} unveiled three distinct shape-dependent-particle behaviors in a confined channel for $\rho /\rho_p=1.05$: (i) the boomerang and triangular particles exhibited an initial large displacement from the centerline toward the channel wall at $y=W$, followed by oscillatory trajectories about the centerline while displaying the largest cumulative angular rotations; (ii) after a large displacement toward the wall at $y=0$, the elliptical and rectangular particles with the same aspect ratio $\left[ (c/b)=(l/w)=2 \right] $ drifted toward the centerline and displayed nearly zero angular rotations as they gradually oriented their principal axis normal to the gravitational field; and (iii) the hexagonal and star particles settled near the centerline similar to a circular particle, but they displayed non-zero cumulative angular rotations, unlike the circular particle. 

As $\rho / \rho_p$ increased from 1.05 to 1.10 (i.e., higher inertial effect), the particles exhibited more oscillations in their settling trajectories as they gradually drifted to the mid-channel. The most striking finding was the effect of the small triangular chip (BDC in Fig. \ref{fig:Geometries}c) on the settling trajectory of the boomerang particle. For $\rho /\rho_p=1.10$,  the small chip was responsible for the persistent periodicity in the boomerang particle's settling trajectory, different from slowly decaying oscillations in the triangular particle's settling trajectory.  Thus, DSP-LBM simulations revealed that a small chip in the boomerang geometry is a key design criteria, controlling the amplitude and frequency of the oscillations in settling trajectories of the boomerang particle.

\begin{figure}[h!]
\begin{center}
\scalebox{0.44} {\includegraphics{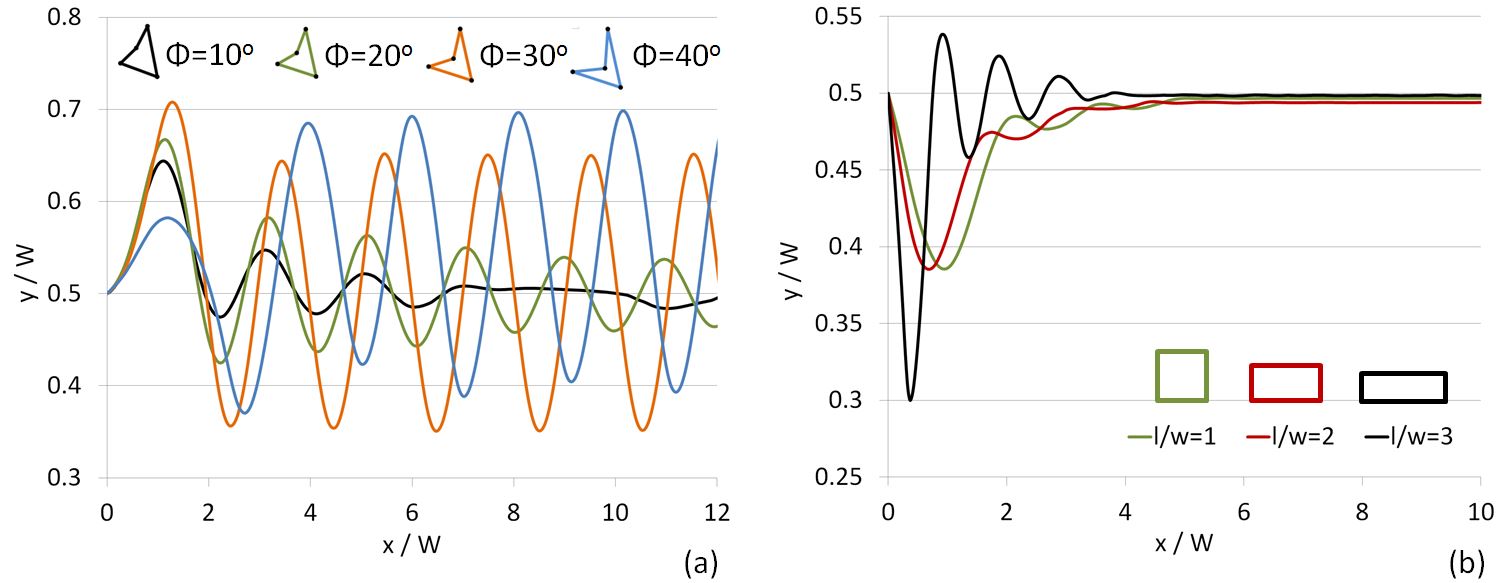} }
\caption{ Settling trajectories of (a) boomerang particles with different surface concavity of its trailing edge (simulations were performed with $\rho/\rho_p=$1.10, $W/R_e$=11.3, $\zeta$=60$^o$ and $A=3.9 \times 10^{-3}$ cm$^2$), and (b) rectangular particles of different aspect ratios (simulations were performed with $\rho/\rho_p=$1.10 and $W/R_e=9.2)$.   \label{Fig:angle_aspect}}
\end{center}
\end{figure}

The other design criteria for engineered DSP may include the (linearized) surface concavity of the boomerang particles and the aspect ratio of rectangular particles. The effect of the surface concavity of the trailing edge of the boomerang particle, controlled by its inner angle ($\phi$) on its settling trajectory, is shown in Fig. \ref{Fig:angle_aspect}a for $\rho/\rho_p=$1.10, $W/R_e$=11.3, $\zeta$=60$^\circ$, and $A_p$=$3.9 \times 10^{-3}$ cm$^2$. DSP-LBM simulations show that the boomerang particle displayed gradually vanishing oscillations in its settling trajectory, similar to the triangular particle, if $\mathbf{x_c}$ was located inside the polygonal surface (for $\phi$=10$^\circ$ and 20$^\circ$). The  boomerang particle exhibited periodic oscillations in its settling trajectory if $\mathbf{x_c}$ was located on the polygonal surface (for $\phi$=30$^\circ$) or outside the polygonal surface (for $\phi$=40$^\circ$). The oscillation frequency,$\vartheta$, dropped from 1.27 s$^{-1}$ to 1.13 s$^{-1}$ as $\mathbf{x_c}$ moved from the polygonal surface (D in Fig. \ref{fig:Geometries}c) to an exterior point outside the polygonal surface. 

The effects of the aspect ratio of a rectangular particle on its settling trajectories are shown in Fig. \ref{Fig:angle_aspect}b for $\rho/\rho_p=$1.10 and $W/R_e$=9.2. Although rectangular particles with different aspect ratios drifted toward the same equilibrium position at the centerline at $x/W \sim6$, the rectangular particle with the largest aspect ratio exhibited the largest initial displacement from the centerline and more frequent and largest oscillations in its settling trajectory, which could be critical in multi-particle flows. 

The effects of particle shape on the settling (terminal) velocity of an individual particle are provided in Supplementary Information-2. As shown in Supplementary Information-3, the overall settling trajectories of DSP in these simulations are deemed to be independent of grid resolution for all practical purposes. 
 
\vspace{0.25 cm}
\noindent
\textbf{Flow Trajectories of Individual Particles of Different Shapes}

\begin{figure}[h!]
\begin{center}
\scalebox{0.44} {\includegraphics{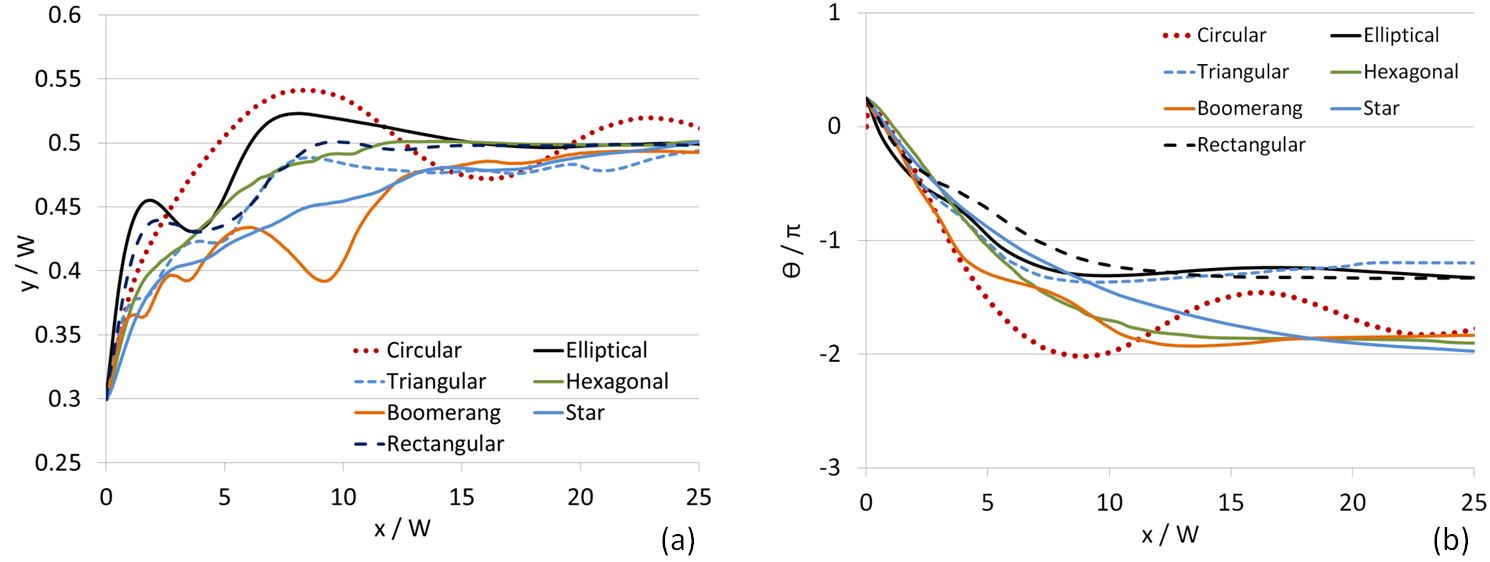} }
\caption{ (a) Flow trajectories and (b) cumulative angular rotations of DSP at $Re=$35. $W/R_e$=11.3 and $\rho_p / \rho =$1.0.   \label{Fig:DSP_flow}}
\end{center}
\end{figure}

In the DSP settling problems discussed above, the fluid was initially quiescent. To simulate shape-dependent flow trajectories of DSP, the particles whose initial orientations shown in Fig. \ref{Fig:Initial_Orientation} were released into a Poiseuille flow from a point 20$\%$ off the centerline after the steady-flow field was established. A neutrally-buoyant spherical particle in a Poiseuille flow typically exhibits the Segre-Silberberg effect\cite{SS62} with an equilibrium settling position between the channel wall and centerline, in which the equilibrium position varies with $Re$ \cite{KGM66,MMG04} (here, $Re=2R U_{ss}/\nu$, where $U_{ss}$ is the average steady fluid velocity prior to releases of the particles). For $Re\sim$0.1 and $W/R_e=6.6$, the equilibrium position of the neutrally-buoyant spherical particle was on the centerline in a tube \cite{KGM66,YWH05}. Consistent with these findings, different equilibrium positions of a circular particle in a Poiseuille flow computed by  DSP-LBM as a function of $Re$ are shown in Supplementary Information-4. Among them, $Re=35.2$, corresponding to the average steady fluid velocity of 4.97 cm/s prior to the particle release, was chosen and the flow trajectories of DSP were simulated (Fig. \ref{Fig:DSP_flow}). At $Re=35.2$, the circular particle exhibited slowly diminishing overshots about the centerline in its flow trajectory due to combined effects of inertial and wall effects. At much higher $Re$, however, the wall effect may be confined to near-wall layers only \cite{A99}.    

When compared to the settling trajectories of DSP (Fig. \ref{Fig:Settling_BC}), the flow trajectories of the DSP are more sensitive to the particle shape in a flowing fluid. DSP followed distinct flow trajectories at $Re=35.2$ before they drifted to their equilibrium position at $x/W\sim$25. Only DCsPs exhibited overshots in their flow trajectories. Although the settling trajectories of the elliptical and rectangular particles with the aspect ratio of 2 were similar, their flow trajectories were different, revealing the significant effect of the (discretized) curved particle surface on particle trajectories in a shear flow. Similarly, the settling trajectories of the star and hexagonal particles were similar, unlike their trajectories in a shear flow. Uniform and repetitive oscillations in the settling trajectories of boomerang and triangular particles were replaced by non-uniform oscillations in their trajectories in a shear flow. Circular, star, hexagonal, and boomerang particles displayed the largest cumulative angular rotations at $Re=35.2$ while the boomerang and triangular particles exhibited the largest cumulative angular rotations as they settled.

\vspace{0.25 cm}
\noindent
\textbf{Settling and Flow of a Mixture of DSP}

The effect of particle shapes on the settling and flow behavior of a mixture of DSP was numerically demonstrated here for the first time. Four simulations were setup, through which trajectories and velocities of seven settling or flowing DSP were compared to those of seven circular particles. All particles, regardless of their shapes, had the same surface area with $R_e=$385 $\mu$m. The interparticle distance at the release location was $4R_e$ and the width and length of the domain was $40R_e \times80R_e$. The fluidic domain was bounded in the settling simulation. A periodic boundary condition was implemented at the inlet and outlet for the flow simulation for which $Re=38$.  Steric interaction forces, based on two-body Lennard-Jones potentials,\cite{BS10} were used to avoid unphysical overlapping of particles when they are in near contact, as described in Supplementary Information-5.   

Figs. \ref{Fig:SevenParticle} a-b show that use of multiple surrogate circular particles in place of a mixture of non-circular particles led to not only misrepresentation of settling trajectories of DSP, but also underestimation of their settling velocities by a factor of up to 2.2 (large velocity ratios near the bottom boundary can be ignored as some particles rested on the bottom while the others continued to roll, which resulted in large velocity ratios). Similarly, Figs. \ref{Fig:SevenParticle} c-d show that if non-circular particle shapes are overlooked, lateral displacements in computed trajectories significantly differed and particle velocities deviated by a factor of $\sim 0.9-1.2$. Accurate displacements and velocities are critical in the design of engineered particles for targeted drug deliveries. Fig. \ref{Fig:SevenParticle} demonstrated that non-circular shapes of particles have pronounced effects on the settling and flow behaviors of a mixture of DSP and their representation by circular shapes introduces errors in calculations of particles trajectories and travel times.

  \begin{figure}[h!]
\begin{center}
\scalebox{0.44} {\includegraphics{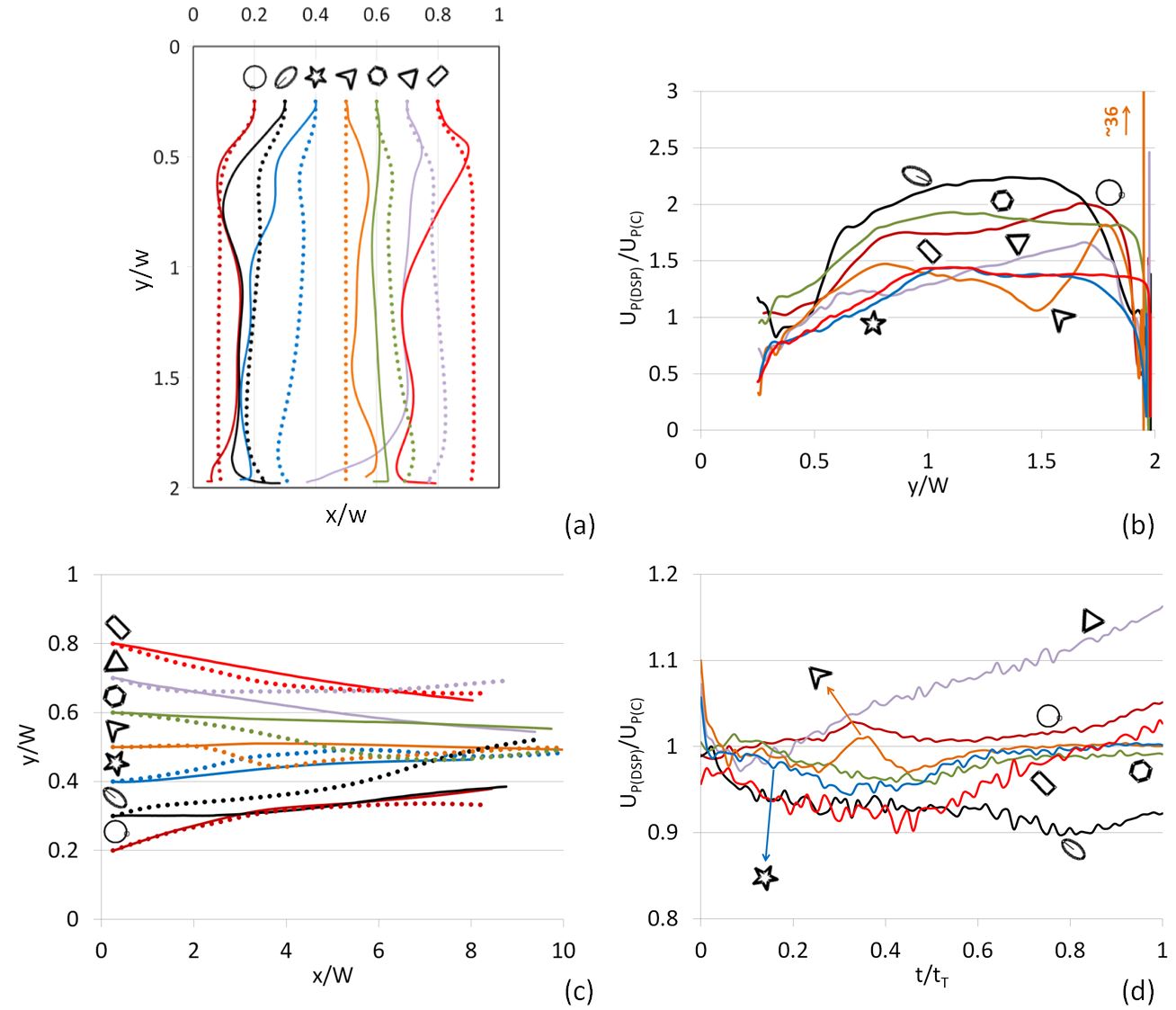} }
\caption{ (a) Comparison of settling trajectories of seven DSP (in solid lines) to seven circular particles (in dashed lines) in an initially quiescent fluid in a confined domain, (b) the ratio of the settling velocity of a different-shape particle to its circular-shape counterpart released from the same point. (c) Comparison of flow trajectories of seven DSP (in solid lines) to seven circular particles (in dashed lines) in a Poiseuille flow with $Re=38$, (d) the ratio of the translational velocity of a different-shaped particle to its circular-shape counterpart released from the same point. $t_T$ is the total simulation time. \label{Fig:SevenParticle}}
\end{center}
\end{figure}

\vspace{0.25 cm}
\noindent

\section*{Discussion}
\begin{figure}[h!]
\begin{center}
\scalebox{0.35} {\includegraphics{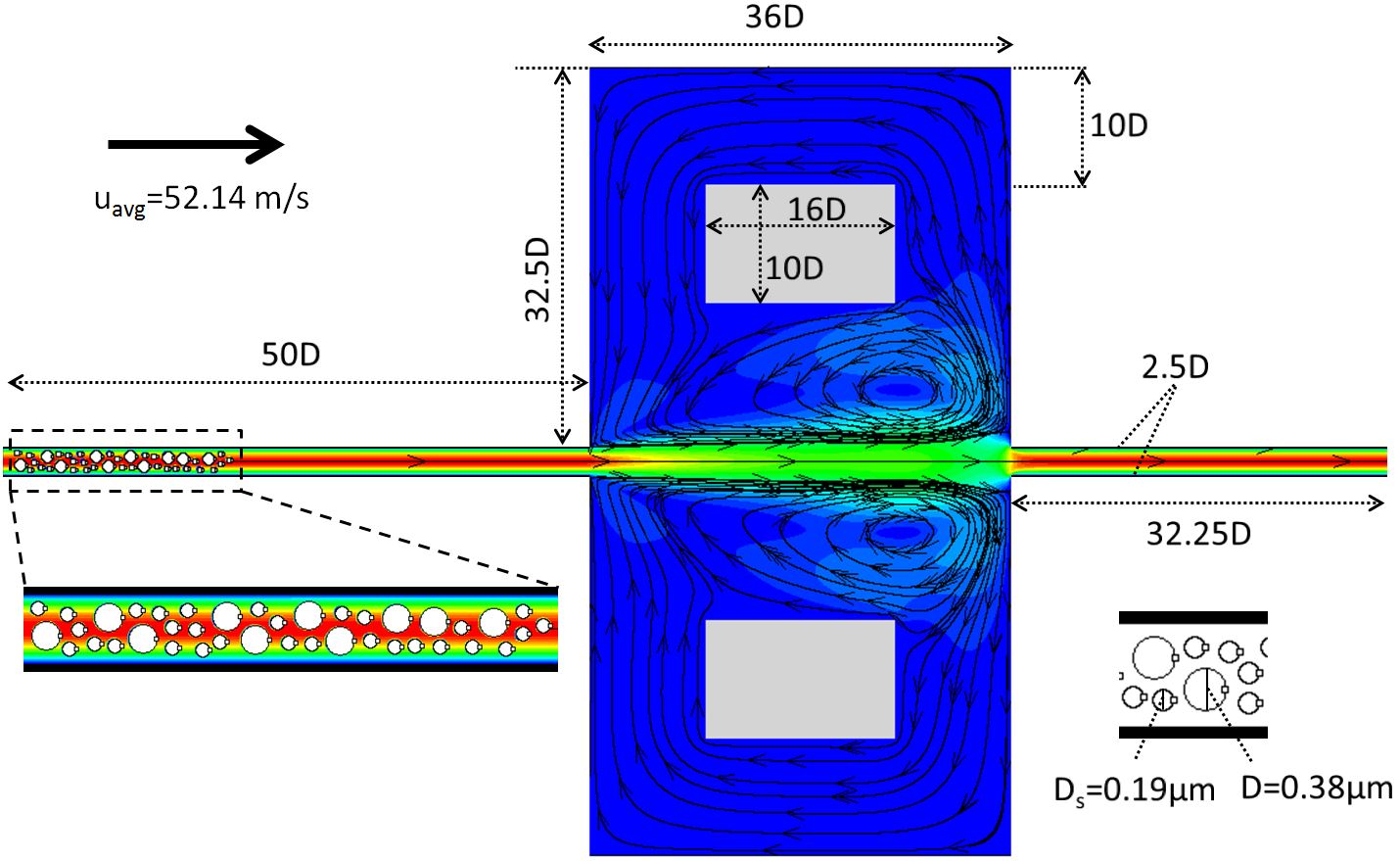} }
\caption{ Steady-flow field in a subsection of microfluidic geometry in Ref \cite{PCD17}. All dimensions are scaled with respect to the large particle diameter, $D$. Small circles attached to particles are used to trace angular rotations of particles.} \label{Fig:flow}
\end{center}
\end{figure}

In the preceding sections, DSP-LBM simulations demonstrated significant effects of particle shapes on the settling or flow trajectories of an individual particle or a mixture of DSP. Using the DSP-LBM, we investigated here the validity of recent findings and implications in microfluidic analyses: (i) would steady vortex structures alone be used to quantify vortex-controlled size-based sorting of particles?  (ii) would larger particles be selectively entrapped in steady vortex regions despite the cumulative effects of particle-fluid hydrodynamics on the fluid velocity in relatively dense suspensions? and (iii) would the findings from vortex-controlled size-based separation of circular particles be extensible to non-circular particles in microfluidics? To answer these questions, DSP-LBM simulations were setup using a single chamber of the microfluidic geometry in Ref. \cite{PCD17} After the steady-flow field was established, 10 large particles of 38$\mu$m in diameter and 30 small particles of 19$\mu$m in diameter were released into a microfluidic chamber from random locations at the inlet. The dimensions of the microfluidic domain and the steady flow field are shown in Fig. \ref{Fig:flow}. The fluid was water with $\nu$=0.01 cm$^2$/s and $c_s$=1,460 m/s, and the particles were neutrally buoyant. The average flow rate, $u_{avg}$, of 52.14 m/s at the inlet in a single-chambered microfluidic chamber produced vortex structures, similar to the vortex structures in a multi-chambered microfluidic device with $u_{avg} \sim$ 1,700 m/s in Fig. 3 of Ref. \cite{PCD17}

\begin{figure}[h!]
\begin{center}
\scalebox{0.43} {\includegraphics{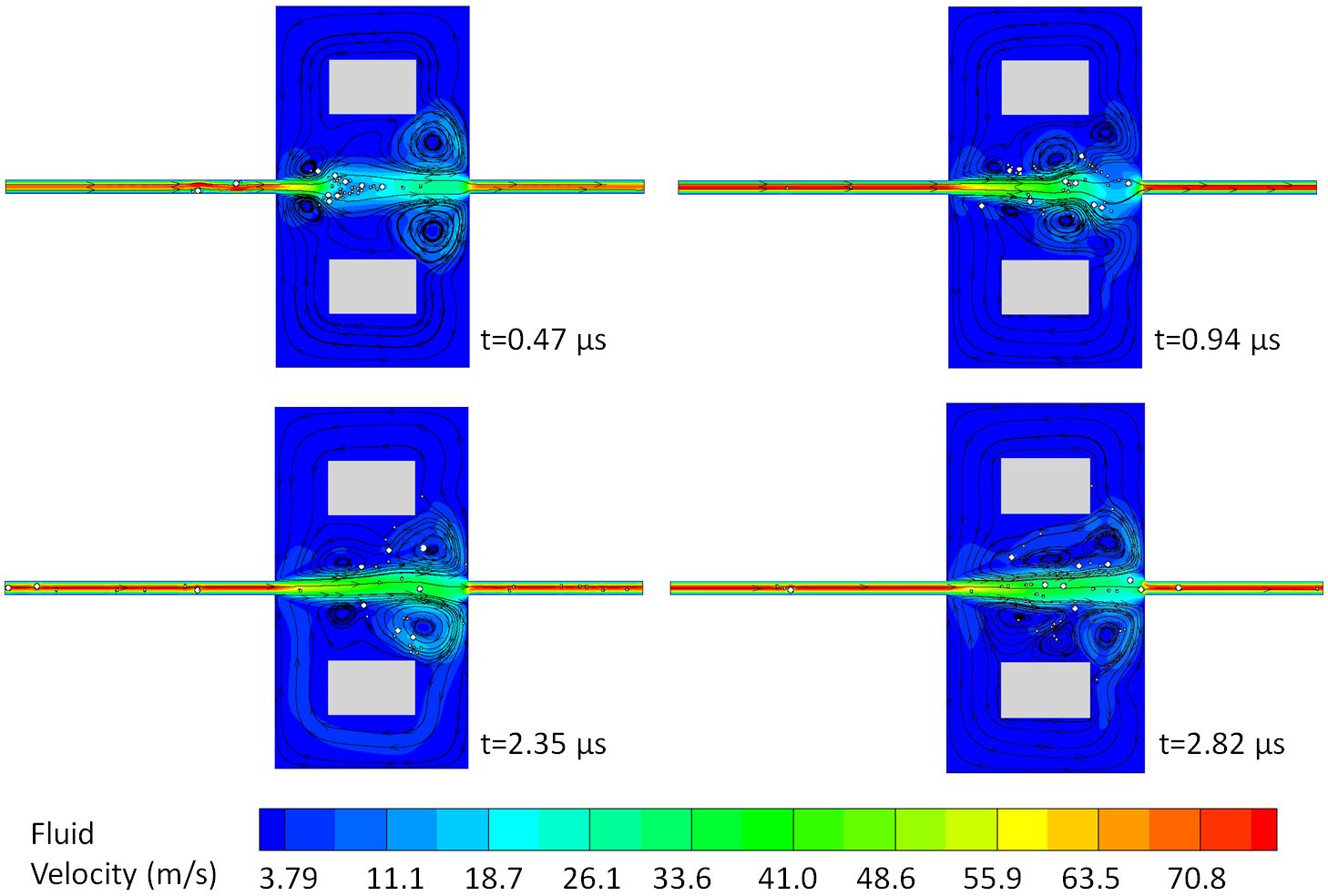} }
\caption{ Transient vortex structures. \label{Fig:vortex_field}}
\end{center}
\end{figure}

Although steady vortex structures were previously envisioned to trap particles in microfluidic devices \cite{ZKP13,PCD17}, Fig. \ref{Fig:vortex_field} shows that vortices in a flowing fluid including mobile particles are indeed unsteady, even if the pressure differential at the inlet and outlet is held constant in time. Symmetry breaks in the flow domain with initially symmetric vortex structures, disappearance or changes in the location of vortices, and formation (birth) of new vortices as a result of cumulative effects of interparticle and particle-fluid hydrodynamics are evident from Fig. \ref{Fig:vortex_field}. Particle motion in this case is largely determined by momentum exchanges between the particles and unsteady discrete vortices, similar to the underlying reasoning of a steadily swimming fish in a water with discrete vortices\cite{HVG15}, for which Lagrangian coherent structures are typically used to decompose unsteady fluid flows into dynamically different regions. In brief, for the initial flow condition given in Fig. \ref{Fig:flow} as in Ref. \cite{PCD17}, the flow field involving multiple mobile particles was inherently transient, which contradicts the use of steady vortex regions \cite{ZKP13,PCD17} in assessing particle entrapments in microfluidics devices. Moreover, the sorting mechanism related to correlations between the lateral displacements of particles to their sizes\cite{SS14} is not applicable for multi-particle simulations in a fluidic domain in Fig. \ref{Fig:flow}.  

%

Next, DSP-LBM simulations were used to investigate if the larger circular particles are selectively trapped in unsteady vortex regions (Fig. \ref{Fig:vortex_field}). Particles leaving the flow domain were allowed to re-enter from the inlet. Simulations continued up to 7.4 $\mu$s, which was long enough for some particles to travel through the entire domain 8 times, referred to as 8 loops here. Flow trajectories of some of the large and small particles are shown in Supplementary Information-6. Table \ref{Table:loop} reports that large circular particles left the flow domain on average 39$\%$ more often than small particles in Fig. \ref{Fig:vortex_field}. Thus, as compared to small particles, large particles had smaller residence times and were less-frequently trapped by transient vortices, different from earlier findings\cite{ZKP13,PCD17} that relied on the assumption of particle entrapments by steady vortices. However, the use of steady vortex structures to assess particle entrapments may still be valid for microfluidics involving dilute suspensions in lower $Re$ flows.

\begin{table}[h]
\caption{Number of trips (loops) the particles experience in a microfluidic device.}
\centering
\begin{tabular}{ |p{3cm}||p{3cm}|p{3cm}|p{3cm}| p{3cm}|  }
\hline
Geometric Shape of LPs* & Total Number of Loops by LPs & Average Number of Loops by a LP & Total Number of Loops by SPs* & Average Number of Loops by a SP \\
 \hline
Circular   & 39  & 3.9 &  84 & 2.8\\
Elliptical & 47  & 4.7 &  54 & 1.8\\
Hexagonal  & 41  & 4.1 &  90 &  3.0\\
  \hline
\end{tabular}
\small{(*) LP stands for large particles of different geometric shapes. SP stands for small circular-cylindrical particles. }
\label{Table:loop}
\end{table}

\begin{figure}[h!]
\begin{center}
\scalebox{0.43} {\includegraphics{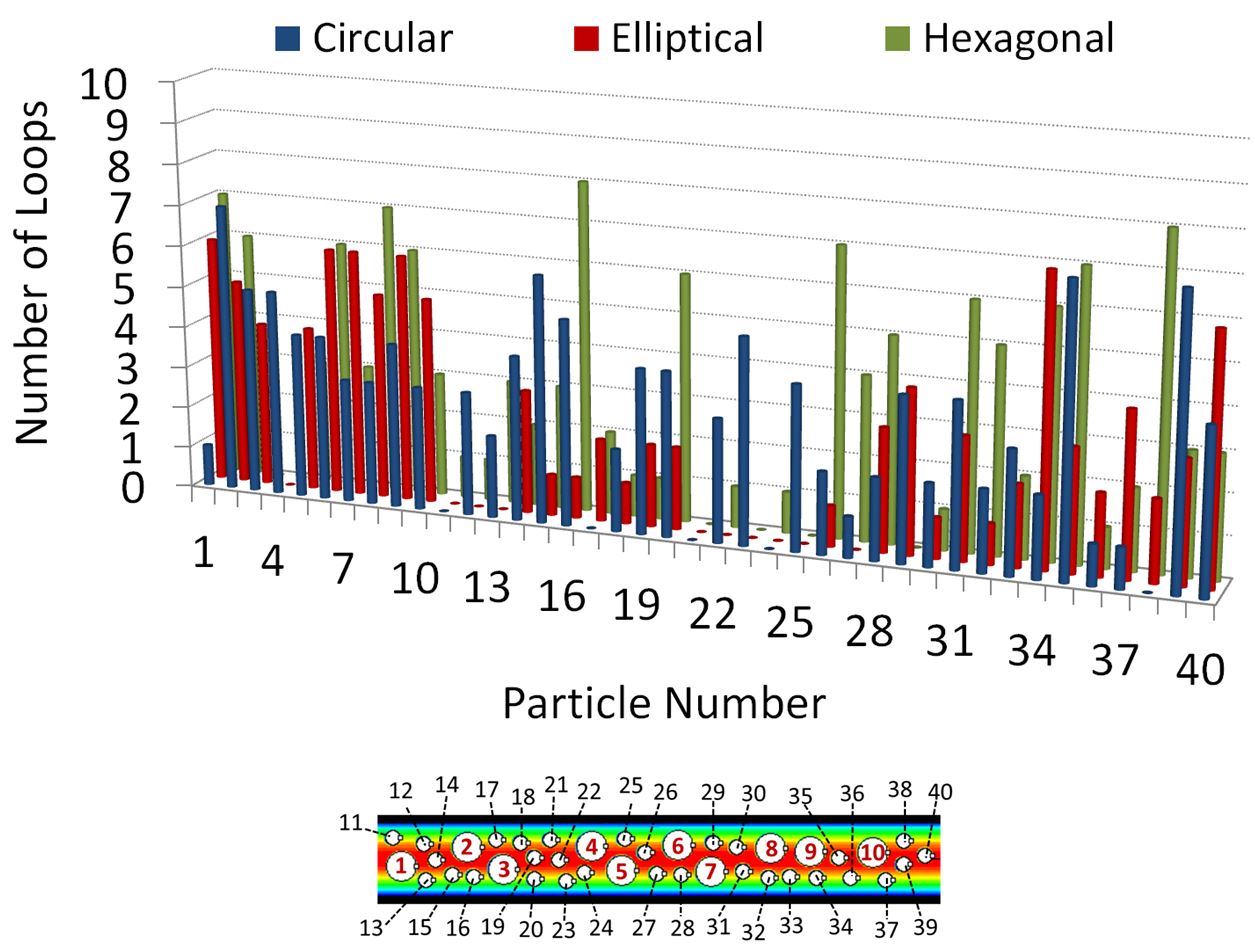} }
\caption{ Number of loops (trips) each particle experienced across the microfluidic domain. \label{Fig:loops}}
\end{center}
\end{figure}

Finally, the effect of the geometric shape of the large particles on the vortex entrapments of small and large particles were investigated for the microfluidic domain in Fig. \ref{Fig:flow}. In DSP-LBM simulations, the shape of the large particles was either circular, elliptical (with an aspect ration of 1.2), or hexagonal with the surface area of 0.11 $\mu$m$^2$, while the small particles were circular. This simulation was setup to mimic a small number of large, non-circular tumor cells dispersed in a large number of small, circular healthy cells. Fig. \ref{Fig:loops} shows that the particle shape affected the residence time of all particles in the microfluidic domain. For example, although Particle 38 was permanently trapped in the microfluidic domain if the large particles were circular, it traveled through the microfluidic domain 8 times if the large particles were hexagonal. Moreover, Table \ref{Table:loop} shows that large hexagonal particles resulted in shorter average residence times for all particles with $5\%$ and $7\%$ increases in the number of loops for small and large particles, respectively. Strikingly, the use of large elliptical particles, instead of large circular particles, resulted in $36\%$ enhanced entrapments for smaller particles, while $21\%$ less entrapments for larger particles. Although these findings require further systematic experimental and numerical analyses to confirm, DSP-LBM simulations showed for the first time that by changing the shape of large particles from circular to elliptical, the smaller particles could be selectively entrapped by transient vortices while the larger particles could be effectively flushed out, which is in contrast to current and proposed uses of microfluidics for vortex-controlled, size-based separation of rigid particles.

In brief, considering strong disparities between flow trajectories of the circular and non-circular particles in microfluidic domains, the use of surrogate spherical particles to mimic tumor cells of abnormal shapes\cite{PAD14,MBL14} in microfluidic experiments as in Ref \cite{PCD17}, could lead to misleading assessments on the performance of the microfluidic designs proposed to isolate CTCs from healthy cells in biofluids. Here, we demonstrated that DSP-LBM could serve as a useful numerical tool for such analyses. 

\section*{Acknowledgements} 

Funding for this research was provided by Southwest Research Institute's Internal Research and Development Program, 18R-8602 and 15R-8651. S.S. wishes to acknowledge funding from the European Research Council under the European Union's Horizon 2020 Framework Programme (No. FP/2014- 2020)/ERC Grant Agreement No. 739964 (COPMAT). The authors thank Miriam R. Juckett of Southwest Research Institute for reviewing the manuscript.  

\section*{Author contributions statement}

H.B. and S.S. developed the numerical model. H.B. ran numerical simulations. D.W. helped with analyzing the results, J.B. involved in coding and simulation runs.  All authors reviewed the manuscript. 

\section*{Additional information}
The authors declare no competing interests. 

\newpage

\section*{Supplementary Information-1: Geometric Description of DSP}

\subsubsection*{Boomerang-shaped Particle} 

The boomerang-shaped particle geometry is described by four vertices $v_{b1}\cdots  v_{b4}$, and three angles, where $\alpha$, $\phi$, and $\zeta$ (Fig. 1c). $\alpha$ is the initial tilt angle of the particle in the counter-clockwise direction, and $\phi$ and $\zeta$ specify the spatial variations of the width of the wings. The particle geometry is generated by carving out a small isosceles triangle, BDC, from a large isosceles triangle, BAC. The locations of vertices of the boomerang-shaped particle are computed by

 \begin{equation} 
  \label{boom_e1} 
 \left[ \begin{array}{c} x_{v_{B1}}\\ y_{v_{B1}} \\ x_{v_{B2}}\\ y_{v_{B2}} \\ x_{v_{B3}}\\ y_{v_{B3}} \\ x_{v_{B4}}\\ y_{v_{B4}} \end{array} \right] = \left[ \begin{array}{c} x_c \\ y_c \\ x_c \\ y_c \\ x_c \\ y_c \\ x_c \\ y_c \end{array} \right] - \frac{B}{4}  
 \left[ \begin{array}{c} 2cos(\alpha) + \eta_c + \varphi_c 
                      \\ 2sin(\alpha) + \eta_s + \varphi_s
                      \\ 2cos(\alpha)+ \eta_c -3 \varphi_c  
                      \\ 2sin(\alpha)+ \eta_s -3 \varphi_s        
                      \\ \eta_c + \varphi_c -6 cos(\alpha) 
                      \\ \eta_s +  \varphi_s -6 sin(\alpha)
                      \\ 2cos(\alpha) + \varphi_c  -3 \eta_c 
                      \\ 2sin(\alpha) + \varphi_s  -3 \eta_s
 \end{array} \right], 
 \end{equation}

\noindent which $\eta_c=cos(\alpha+\zeta) / cos (\zeta)$, $\varphi_c=cos(\alpha+\phi)/cos(\phi)$, $\eta_s=sin(\alpha+\zeta) / cos (\zeta)$, and $\varphi_s=sin(\alpha+\phi)/cos(\phi)$. The mass of the boomerang-shaped particle per unit particle thickness is given by $m_p= A_B \rho_p$, in which the surface area of the boomerang-shaped geometry is $A_B=B^2 \left[ tan (\zeta) - tan (\phi)  \right]$ and $B$ is half of the base length of the triangle BDC. The particle's moment of inertia is computed via $I_p= \frac{m_p}{72} \left[ 4 \left( H^2+Hh+h^2 \right) + 3B^2  \right] -\kappa^2 A_{BDC} $, where $H$ and $h$ are the heights of the triangle BAC and BDC, in which $H=B tan(\zeta)$ and $h=B tan(\phi)$, $\kappa$ is the distance between the center of mass of the triangle BAC and triangle BDC, and $A_{BDC}$ is the area of the triangle BDC.  

\subsubsection*{Equilateral Triangular-shaped Particle} 

The equilateral triangular particle geometry is represented by the side length of $a$, three vertices, $v_{T1}-v_{T3}$, and an initial tilt angle, $\alpha$, in the counter-clockwise direction (Fig. 1d). The coordinates of the vertices of the triangular particle are given by

 \begin{equation} 
  \label{tr_e1} 
 \left[ \begin{array}{c} x_{v_{T1}}\\ y_{v_{T1}} \\ x_{v_{T2}}\\ y_{v_{T2}} \\ x_{v_{T3}}\\ y_{v_{T3}} \end{array} \right] =\left[ \begin{array}{c} x_c \\ y_c \\x_c \\ y_c \\x_c \\ y_c \end{array} \right]+ \frac{2h}{3}  \left[ \begin{array}{c} cos\left( \pi / 2 + \alpha \right) \\  sin \left( \pi / 2 + \alpha \right)  \\ - sin \left( \pi / 3 - \alpha \right)\\   - cos \left( \pi / 3 - \alpha \right) \\  sin \left( \pi / 3+ \alpha \right) \\ cos \left( \pi / 3 + \alpha \right) \end{array} \right], 
 \end{equation}

 \noindent
in which $h$ is the height of a triangle, $h=\frac{\sqrt{3}}{2} a$. The particle mass per unit thickness is given by $m_p=A_T \rho_p$, in which the surface area of the  equilateral triangle is $A_T=\frac{\sqrt{3}}{4} a^2$. The moment of inertia of a triangular particle is computed from $I_T=\frac{m_p}{72} \left( 3a^2+4h^2 \right)$.

\subsubsection*{Hexagonal-shaped Particle} 

The hexagonal particle geometry is described by an uniform side length of $L$ and initial tilt angle, $\alpha$ (Fig. 1e). The locations of vertices of the hexagonal particle, $v_{H1}\cdots  v_{H6}$, are computed by

 \begin{equation} 
  \label{hp_e1} 
 \left[ \begin{array}{c} x_{v_{Hi}}\\ y_{v_{Hi}} \end{array} \right] =\left[ \begin{array}{c} x_c \\ y_c \end{array} \right]+   L  \left[ \begin{array}{c} cos \left(\alpha+\left(i-1 \right) \pi /3 \right) \\  sin\left(\alpha+\left( i-1 \right) \pi /3\right) \end{array} \right], 
 \end{equation}

\noindent
in which $i \epsilon \left[1,6\right]$. The mass of a hexagonal-shaped particle per unit particle thickness is $m_p=A_H \rho_p$, in which the surface area of the particle is $A_H =3\sqrt2 L^2 /2$.  The moment of inertia of a hexagonal particle is $I_p = \frac{m_p L^2}{24} \left[ 1+3cot^2 \left(  \frac{\pi}{6}  \right)  \right]$.

\subsubsection*{Rectangular-shaped Particle} 

The rectangular-shaped particle geometry is represented by two side lengths of $l$ and $w$, four vertices, $v_{R1} - v_{R4}$, and an initial tilt angle, $\alpha$, in the counter-clockwise direction  (Fig. 1f). The coordinates of the vertices are 

 \begin{equation} 
  \label{rec_e1} 
 \left[ \begin{array}{c} x_{v_{R1}}\\ y_{v_{R1}} \\ x_{v_{R2}}\\ y_{v_{R2}} \\ x_{v_{R3}}\\ y_{v_{R3}} \\ x_{v_{R4}}\\ y_{v_{R4}}   \end{array} \right] =\left[ \begin{array}{c} x_c \\ y_c \\ x_c \\ y_c \\ x_c \\ y_c \\ x_c \\ y_c \end{array} \right]+  \frac{ \sqrt{l^2+w^2} }{2}  \left[ \begin{array}{c} cos \left(\alpha +\theta \right)  \\  sin \left(\alpha +\theta \right) \\ -cos \left(\alpha -\theta \right)  \\  sin \left(\alpha -\theta \right)  \\ -cos \left(\alpha +\theta \right)  \\  -sin \left(\alpha +\theta \right)  \\ cos \left(\alpha -\theta \right)  \\  -sin \left(\alpha -\theta \right)         \end{array} \right]. 
 \end{equation}

 \noindent
The particle mass per unit thickness was given by $m_p=A_R \rho_p$, in which the surface area of the  rectangular geometry is $A_R=lw$. The moment of inertia of a rectangular particle is computed from $I_R=\frac{m_p}{12} \left( l^2+h^2 \right)$. 

\subsubsection*{Circular-shaped Particle} 

The circular particle geometry is constructed by equally-spaced boundary nodes, $N_{bnd}$, along its curved surface and the particle radius, $R$ (Supplementary Fig. \ref{Fig:Circular}). The coordinates of the boundary nodes (denoted by small circles in Supplementary Fig. \ref{Fig:Circular}) are computed by

 \begin{equation} 
  \label{cir_e1} 
 \left[ \begin{array}{c} x_i\\ y_i \end{array} \right] = \left[ \begin{array}{c} x_c \\ y_c \end{array} \right]+R \left[ \begin{array}{c} cos\left( 2\pi \left( i-1 \right) / \left(N_{bnd}-1 \right)  \right) \\ sin \left( 2 \pi \left( i-1 \right)  / \left( N_{bnd}-1 \right)  \right) \end{array} \right]. 
 \end{equation}

 \noindent
The mass of the circular particle per unit particle thickness is $m_p=A_C \rho_p$, in which the surface area of the circular particle is $A_C=\pi R^2$. When the circular particle is treated as a thin solid disk, its moment of inertia is computed via $I_C=\frac{1}{2}m_p {R}^2$.

\begin{suppfigure}[h!]
\begin{center}
\scalebox{0.37} {\includegraphics{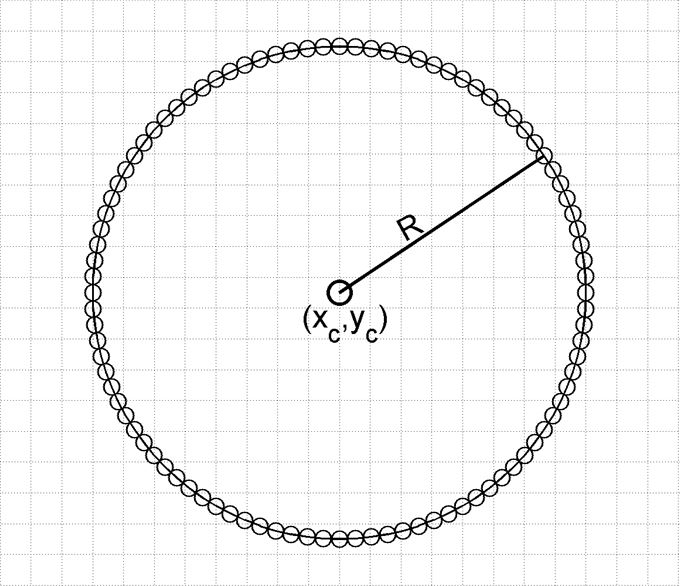}} 
\caption{A schematic representation of a circular particle geometry in the LBM.} \label{Fig:Circular}
\end{center}
\end{suppfigure}

\newpage

\section*{Supplementary Information-2: Settling (Terminal) Velocities of DSP}

The settling velocity of a particle approaches a constant (equilibrium) value as the buoyant force is balanced by the viscous drag force  \cite{MM02} . The DSP-LBM was used to calculate the effect of particle shapes on the settling (terminal) velocities, $U_s$, of DSP, as a function of particle density and the blockage ratio, $W/R_e$, in which $W$ is the channel width and $R_e$ is the effective particle radius as described in the main text. From a series of experiments conducted with glass spheres ranging in size from 0.1 $\mu$m to 6 {mm} in diameter, an empirical relation (Eq. \ref{e.SV_Sp}) was constructed by Gibbs \textit{et al.} \cite{GML71} to determine the settling velocity of spherical particles in an initially quiescent water in a bounded domain,

\begin{equation}
 \label{e.SV_Sp}
U_s=\frac{-3\rho \nu + \sqrt[]{9\left( \rho \nu \right)^2+|\mathbf{g}|R_p^2 \rho \left( \rho_p - \rho \right) \left( 0.015476+0.19841R_p \right) }} {\rho \left(  0.011607+1.14881R_p  \right) }.
 \end{equation}

 \begin{suppfigure}[h!]
\begin{center}
\scalebox{0.30} {\includegraphics{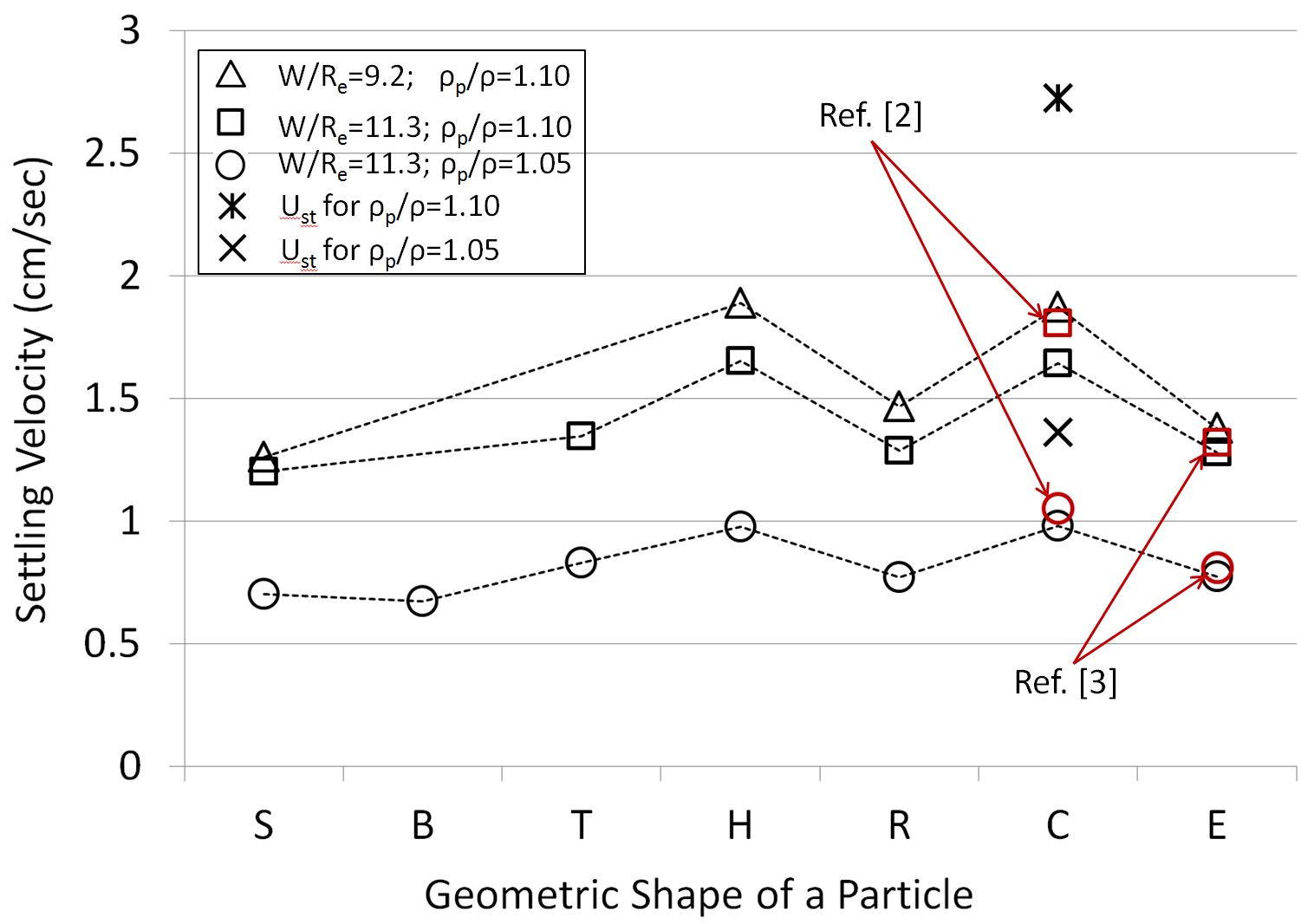} }
\caption{ DSP-LBM simulations of settling (terminal) velocities, $U_s$, of DSP with different densities and blockage ratios. The letters in the horizontal axis corresponds to the first letter of particle geometries. Red symbols correspond to the literature data.   \label{Fig:Terminal_Vel}}
\end{center}
\end{suppfigure}

\noindent  $U_s$ computed by Eq. \ref{e.SV_Sp} for a spherical particle with $\rho_p/\rho=1.05$ and $1.1$ is in a good agreement with the settling velocities of a circular particle with $\rho_p/\rho=$1.05 and $\rho_p/\rho=1.10$ computed by the DSP-LBM in Supplementary Fig. \ref{Fig:Terminal_Vel}.  Moreover, the settling velocities of an elliptical particle for $\rho_p/\rho=1.05$ and $1.1$ computed by the DSP-LBM are in good agreement with particle settling velocities reported by Xia \textit{et al.} \cite{ZCR09}.  

Supplementary Fig. \ref{Fig:Terminal_Vel}  also reports $U_s$ for the circular particle computed by the Stokes equation, given by $U_{st}={ \left( \rho_p - \rho \right) } \left( 2R_p\right)^2  / {18 \left(\rho \nu \right) }$. $U_{st}$ overestimated $U_s$ computed by the DSP-LBM and Eq. \ref{e.SV_Sp} by 25$\%$ for $\rho_p / \rho=$1.05 and 47$\%$ for $\rho / \rho_p=$1.10, and hence, the settling of the particle with $\rho_p / \rho=$1.05 and $\rho_p / \rho=$1.10 cannot be correctly expressed by Stokes equation. The results shown in Fig. \ref{Fig:Terminal_Vel} also revealed that the settling velocities of the circular, rectangular, and hexagonal particles were more sensitive to the blockage ratio than those of star and elliptical particles.

\newpage

\section*{Supplementary Information-3: Grid Resolution}

To demonstrate the insensitivity of the settling velocities and trajectories of DSP to grid resolution, grid resolution was doubled (2x-Resolution) such that the surface area of each particle was represented by 530 lattice cells, as compared to 265 lattice cells for the base case (x-Resolution) in DSP-LBM simulations. The resultant discrepancies in the settling velocities of DSP were within 2$\%$  (Supplementary Table \ref{Table1}).

\begin{supptable}[h!]
	\caption{Settling velocities of DSP (in cm/s) for $\rho_p / \rho=1.05$ at $x/W=20$, as a function of a grid resolution.} \label{tab:GR}
	\centering
	\begin{tabular}{c|c|c|c|c|c|c|c}
		\hline
		\     & Star & Boomerang & Triangular & Hexagonal & Rectangular & Circular& Elliptical\\
		\hline
		
		x-Resolution          &	0.00900  & 0.00865 & 0.01064 & 0.01250 & 0.00987 &  0.01254 & 0.00990\\ 
		2x-Resolution         &	0.00899  & 0.00856 & 0.01051 & 0.01243 & 0.01007 &  0.01268 & 0.00998\\
\hline
       $\%$ discrepancy     &	0.15     & 1.07    & 1.26    &  0.58  &   1.94   &  1.10    & 0.82\\
		\hline
	\end{tabular}
	\small{
		
	}
	\label{Table1}
\end{supptable}

The trajectories of the boomerang, triangular, elliptical, and rectangular particles with $\rho_p / \rho=1.05$ for x- and 2x-Resolutions are shown in Supplementary Fig. \ref{Fig:Grid_Res1}. For $\rho_p / \rho=1.05$, star-shaped and hexagonal shaped particles settled near the centerline, but their settling trajectories exhibited relatively larger lateral displacements when $\rho_p / \rho=1.10$. Therefore, the sensitivity of their trajectories to grid resolution is shown for $\rho_p / \rho=1.10$ in Supplementary Fig. \ref{Fig:Grid_Res2}.

 \begin{suppfigure}[h]
\begin{center}
\scalebox{0.40} {\includegraphics{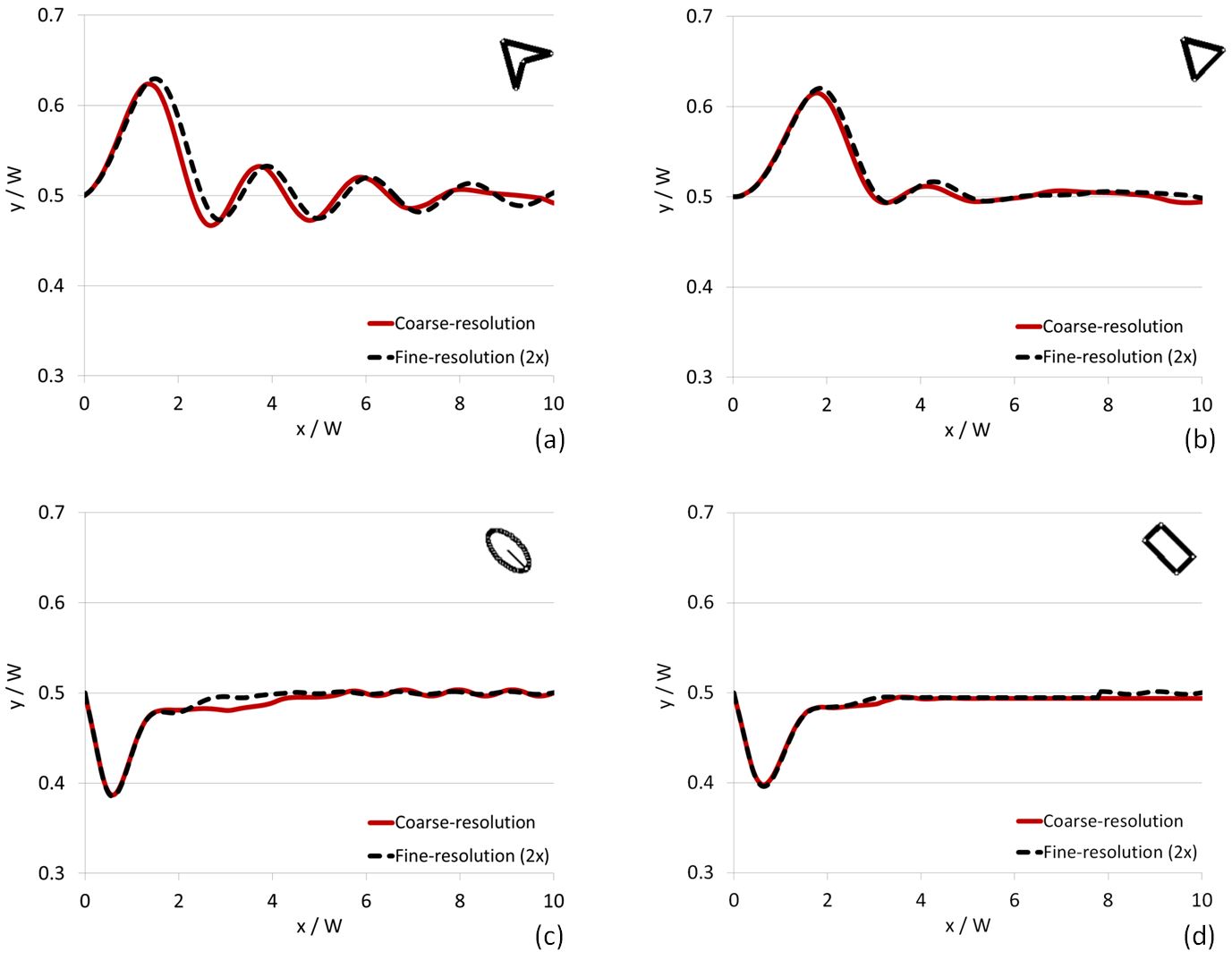} }
\caption{ The effect of grid resolution on the trajectories of a (a) boomerang, (b) triangular, (c) elliptical and (d) rectangular particle for $\rho_p / \rho=1.05$.   \label{Fig:Grid_Res1}}
\end{center}
\end{suppfigure}

 \begin{suppfigure}[h!]
\begin{center}
\scalebox{0.40} {\includegraphics{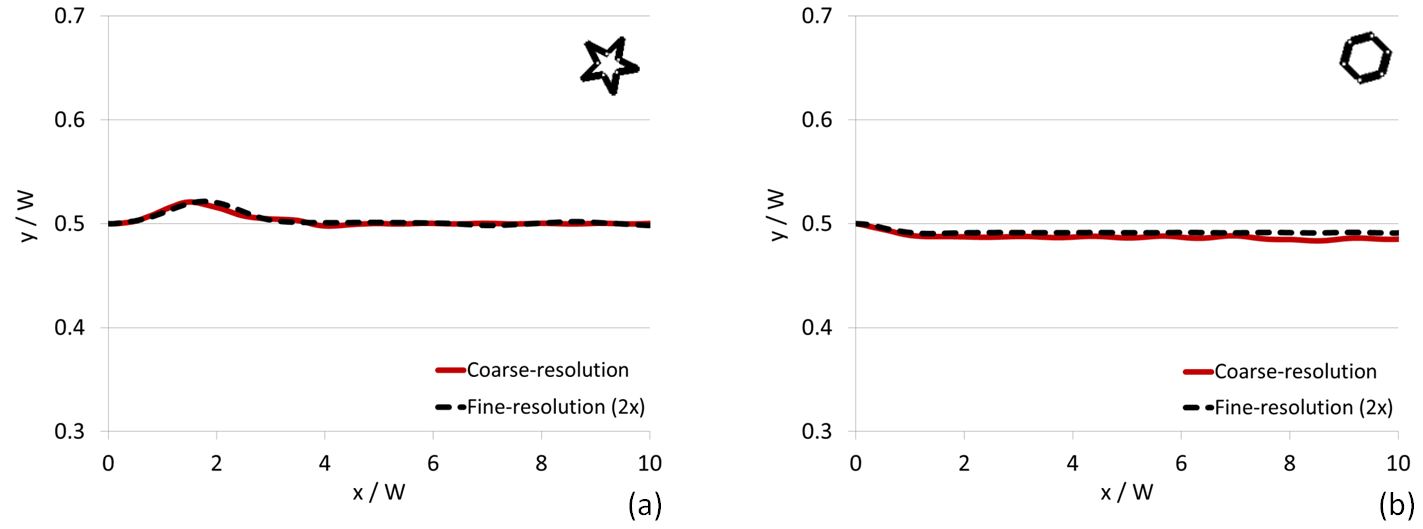} }
\caption{ The effect of grid resolution on the trajectories of a (a) star and (b) hexagonal particle for $\rho_p / \rho=1.10$.   \label{Fig:Grid_Res2}}
\end{center}
\end{suppfigure}

\newpage

In brief, the general trend of settling trajectories of DSP at the relatively coarser and  two-times finer resolutions were similar, as shown in Supplementary Figs. \ref{Fig:Grid_Res1} and \ref{Fig:Grid_Res2}. Moreover, the settling velocities of DSP at the finer and coarser resolutions differed only by $ \le 2\%$ (Supplementary Table \ref{tab:GR}). Therefore, the coarser grid resolution was deemed sufficient for numerical simulations in this paper.


\newpage

\section*{Supplementary Information-4: Flow Trajectories of a Circular-Cylindrical Particle in a Poiseuille Flow}

Flow trajectories of a neutrally-buoyant circular particle in a Poiseuille flow at different $Re$ are shown in Supplementary Fig. \ref{Fig:Flow_tr_sphere}. In these simulations,  $2R_p=$ 0.08 cm, $\nu=$0.01 cm$^2/$s, and $W/R_p$ = 11.3.

 \begin{suppfigure}[h!]
\begin{center}
\scalebox{0.30} {\includegraphics{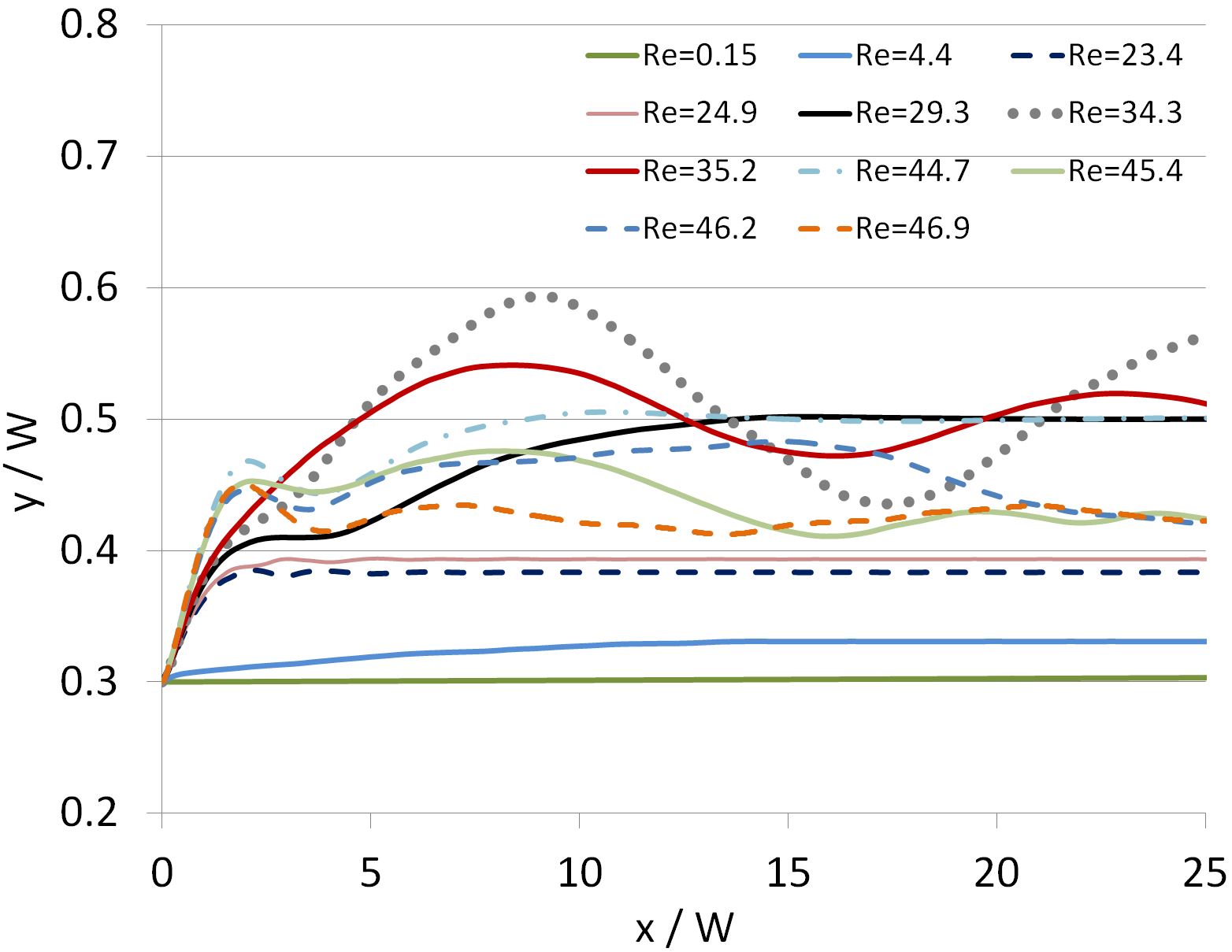} }
\caption{ Flow trajectories of a neutrally-buoyant circular particle in Poiseuille flow at different $Re$.   \label{Fig:Flow_tr_sphere}}
\end{center}
\end{suppfigure}

\newpage
\section*{Supplementary Information-5: Particle-Particle and Particle-Wall Steric Interaction Forces }

Steric interaction forces, $\mathbf{F}_{\mathbf{r}_{i}}$, between the particles and between the particles and stationary solid zones, including channel walls and inline obstacles, are expressed in terms of two-body Lennard-Jones potentials \cite{BS10} such that $\mathbf{F}_{\mathbf{r}_{i}}=-\psi \left( \frac{\mid \mathbf{r}_i \mid} {\mid \mathbf{r_{it}} \mid} \right)^{-13} \mathbf{n}$, where $\mid \mathbf{r}_i \mid$ is the distance between a particle surface node and the neighboring particle surface node ($\mathbf{r}_{i} = \mathbf{r}_{pp'}$) or between a particle surface node and the stationary solid node located on channel walls or inline obstacles ($\mathbf{r}_{i} = \mathbf{r}_{pw}$); $p$ is the particle index; $\mid \mathbf{r}_{it} \mid$ is the repulsive threshold distance; $\mathbf{n}$ is the unit vector along $\mathbf{r}_{i}$; and $\psi$ is the stiffness parameter used to adjust the repulsive strength between the particles and between the particles and stationary solid zones. The total particle-fluid hydrodynamic forces are computed by

\begin{equation}
 \label{e.6}
 \mathbf{F}_T=\sum_{\mathbf{r}_b}
\mathbf{F}_{\mathbf{r}_b}+\sum_{\mathbf{r}^{c,u}_b}\mathbf{F}_{\mathbf{r}^{c,u}_b}
+\sum_{\mid \mathbf{r}_{pw} \mid \le \mid \mathbf{r}_{it} \mid} \mathbf{F}_{\mathbf{r}_{pw}}
+\sum_{\mid \mathbf{r}_{pp'} \mid \le \mid \mathbf{r}_{it} \mid} \mathbf{F}_{\mathbf{r}_{pp'}},
 \end{equation}
 
 \noindent in which the first term on the right-hand-side of Eq. \ref{e.6} is used to calculate particle-fluid hydrodynamic forces at the surface (boundary) nodes located at $\mathbf{r}_b$, the second term is used to calculate forces associated with the uncovered or covered lattice nodes at ${\mathbf{r}^{c,u}_b}$ due to particle motion, the third term is used to calculate interparticle steric interaction forces, and the fourth term is used to calculate the steric interaction forces between the particles and stationary solid zones. Please see Ref. \cite{BS10} for a more detailed explanation of these terms.
 
In this formulation, the total steric interaction forces on a particle exerted by neighboring particles scale with the number of surface nodes of neighboring particles enclosed by an envelope $\mid \mathbf{r}_{it} \mid$ around the particle of interest. The same scaling is also applicable for the steric interactions between the particles and stationary solid zones. In our simulations, $\mid \mathbf{r}_{it} \mid=2.5$ lattice unit (l.u.) away from particle surface and $\psi=1$. Interparticle steric interaction forces are non-zero only when surface boundary nodes of neighboring particles are within 2.5 l.u. of the particle of interest to avoid physically unrealistic overlaps. 

Let the total particle-fluid hydrodynamic force, not excluding the interparticle and particle-wall steric interaction forces, be  $\mathbf{F}_H=\sum_{\mathbf{r}_b} \mathbf{F}_{\mathbf{r}_b}+\sum_{\mathbf{r}^{c,u}_b}\mathbf{F}_{\mathbf{r}^{c,u}_b}$ and interparticle steric interaction forces be $ \mathbf{F}_{PP}=\sum_{\mid \mathbf{r}_{pp'} \mid \le \mid \mathbf{r}_{it} \mid} \mathbf{F}_{\mathbf{r}_{pp'}}$. In our simulations, $\mid \mathbf{F}_H \mid$ is typically on the order of $\mid \mathbf{F}_{pp} \mid$, if the surface (boundary) nodes of neighboring particles are separated from boundary nodes of the particle of interest by a distance $\mid \mathbf{d} \mid$ such that $1$ l.u. $\le \mid \mathbf{d} \mid \le 2.5$ l.u.  However, if  $ \mid \mathbf{d} \mid < 1.0$, $\mathbf{F}_{\mathbf{r}_{pp}}$ results in an instantaneous, short-lived relatively large steric pulse to keep the separation distance larger than 1 l.u. (Fig. \ref{Fig:LJ}).  The movie file showing the settling of a mixture of DSP in Figs. 8a-b, using the interparticle and particle-wall steric interactions in Eq. \ref{e.6}, is provided as a Supplementary Movie file.    

 \begin{suppfigure}[h!]
\begin{center}
\scalebox{0.25} {\includegraphics{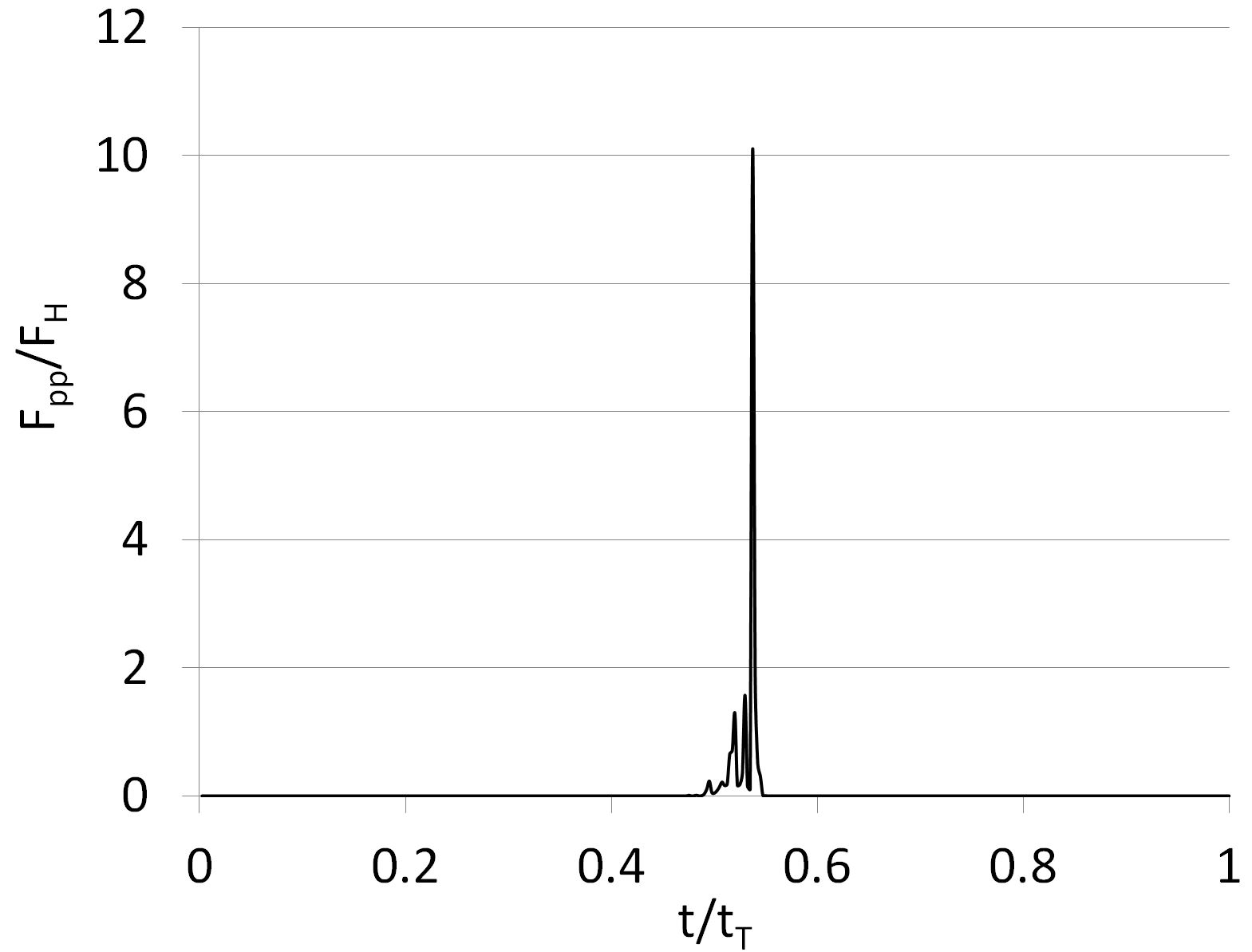} }
\caption{The ratio of the interparticle steric interaction forces to the total particle-fluid hydrodynamics forces, excluding steric interaction forces, acting on the settling elliptical particle. $t_T$ is the total simulation time. \label{Fig:LJ}}
\end{center}
\end{suppfigure}

\section*{Supplementary Information-6: Particles Trajectories in a Microfluidic Device }
Particles of different shapes and sizes were released into water after the steady flow field was established in a microfluidic device geometry in Fig. 9. Transient flow trajectories of four of these particles are shown in Supplementary Fig. \ref{Fig:Trajectories_in_Microfluidic}, which reveals that neither large particles nor small particles were permanently trapped in steady vortex structures established in the fluid prior to releases of the particles. 

The particle L1 (L represents the large particles) was trapped temporarily in the upper and lower halves during its first trip, but eventually escaped the entrapments and left the flow domain. In its second trip, it was trapped in the lower half. Unlike the particle L1, the particle L10, completed the first three trips without being trapped but got trapped in the lower half in its fourth trip. The particle S12 (S represents small particles) completed its first trip with relatively short-lived entrapments, avoided entrapments in its second and third trips, but displayed a prolonged entrapment in its fourth trip. The particle S22, on the other hand, exhibited entrapments in its first and third trips, but flew smoothly in its second trip without any entrapments. 

Numerical simulations revealed similar flow behaviors for all 40 particles. None of these particles were permanently entrapped in transient vortices or in steady vortex regions. The flow field was inherently transient after releases of particles into an initially steady flow field in Fig. 9. The location, size, and number of vortex structures continuously altered as the mobile particles continuously exchanged momentum with the fluid. Hence, the steady vortex structures in Fig. 9 are not responsible for particle entrapments in this simulation.

 \begin{suppfigure}[h!]
\begin{center}
\scalebox{0.38} {\includegraphics{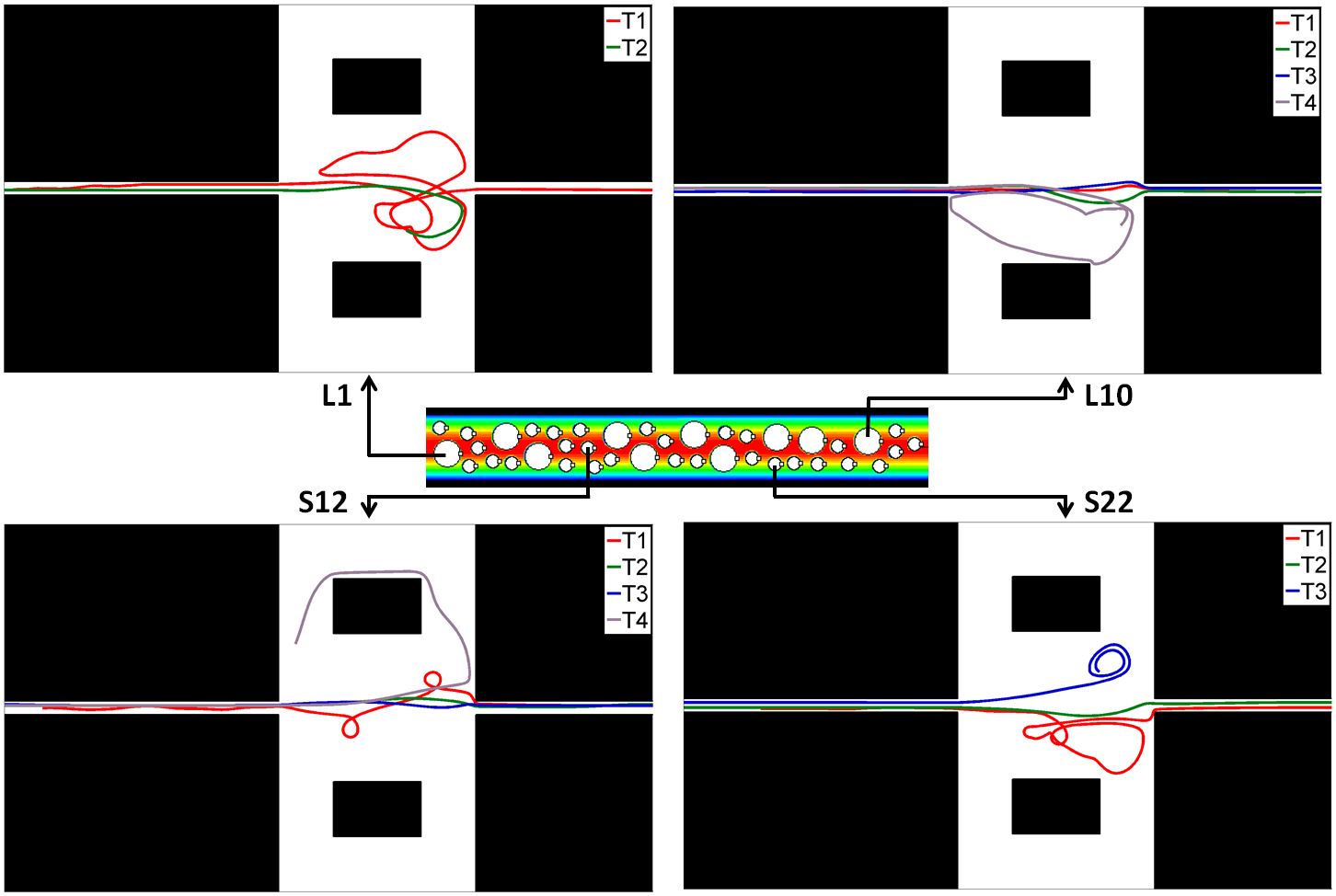} }
\caption{ Flow trajectories of four particles in a microfluidic chamber after the steady flow field was established. T(i) refers to the (i)th trip of the particle in the microfluidic chamber. L denotes large particles and S denotes small particles. Particles leaving the domain from the exit-end were allowed to re-enter the flow domain from the inlet.  \label{Fig:Trajectories_in_Microfluidic}}
\end{center}
\end{suppfigure}

\newpage


\end{document}